\documentclass[12pt,letterpaper,twopage,pdf]{article}
\pdfoutput=1
\usepackage{latexsym}
\usepackage[latin1]{inputenc}   
\usepackage{amsmath}
\usepackage{epsfig}
\usepackage{graphics}
\usepackage{amsfonts}
\usepackage{pifont}
\usepackage{booktabs}
\usepackage{xspace}
\usepackage{color}
\usepackage{booktabs}
\usepackage{xtab}
\usepackage{eurosym}
\usepackage{makeidx}
\usepackage{fancyhdr}
\usepackage{lineno}
\usepackage{charter}

\definecolor{lgrey}{rgb}{0.4,0.4,0.4}            

\newcounter{bibnumber}
\newcounter{impnumber}
\usepackage{feynmp}
\usepackage{xspace}
\usepackage{booktabs}
\usepackage[small, nooneline, center]{subfig}
\usepackage{amsmath,amssymb,bm}
\usepackage{longtable}
\usepackage[colorlinks=true, urlcolor=blue, linkcolor=blue, citecolor=blue]{hyperref}
\usepackage{mciteplus}
\definecolor{lightgray}{rgb}{0.85,0.85,0.87}
\definecolor{gray}{rgb}{0.5,0.5,0.5}
\definecolor{darkgray}{rgb}{0.36,0.36,0.36}
\def\lsim{\mathrel{\rlap{\lower4pt\hbox{\hskip1pt$\sim$}}
    \raise1pt\hbox{$<$}}}                
\def\gsim{\mathrel{\rlap{\lower4pt\hbox{\hskip1pt$\sim$}}
    \raise1pt\hbox{$>$}}}                

\newlength{\tabcolsepsave}

\makeindex

\hypersetup{pdfauthor={Stefano Profumo},pdftitle={Astrophysical Probes of Dark Matter}}
\usepackage[top=1.135in,bottom=1.17in,left=1.16in,right=1.16in]{geometry}
\begin{document}
\title{TASI 2012 Lectures on\\ Astrophysical Probes of Dark Matter\thanks
                 {Lecture Notes for TASI 2012: Theoretical Advanced
 Study Institute in Elementary Particle Physics  --- Searching for
 New Physics at Small and Large Scales. University of Colorado, Boulder,
 CO, June 4 -- 29, 2012.}}
\author{Stefano Profumo\thanks
                 {profumo@ucsc.edu}\\[4mm]
 Department of Physics and Santa Cruz Institute for Particle Physics\\ University of California, Santa Cruz, CA 95064, United States of America}

\date{June, 2012}
\maketitle
\begin{abstract}
\noindent What is the connection between how the dark matter was produced in the early universe and how we can detect it today? Where does the WIMP miracle come from, and is it really a ``WIMP'' miracle? What brackets the mass range for thermal relics? Where does $\langle\sigma v\rangle$ come from, and what does it mean? What is the difference between chemical and kinetic decoupling? Why do some people think that dark matter cannot be lighter than 40 GeV? Why is $b \bar b$ such a popular annihilation final state? Why is antimatter a good way to look for dark matter? Why should the cosmic-ray positron fraction decline with energy, and why does it not? How does one calculate the flux of neutrinos from dark matter annihilation in a celestial body, and when is that flux independent of the dark matter pair-annihilation rate? How does dark matter produce photons? --- Read these lecture notes, do the suggested 10 exercises, and you will find answers to all of these questions (and to many more on what You Always Wanted to Know About Dark Matter But Were Afraid to Ask).
\end{abstract}
\clearpage

\tableofcontents
\vspace*{0.8cm}

\hrule\vspace{5mm}

\section*{Introduction \label{sec:introduction}}
\addcontentsline{toc}{section}{Introduction}

These are lecture notes for a series of four lectures on astrophysical probes of dark matter I delivered in June 2012 at the Theoretical Advanced Study Institute in Elementary Particle Physics (TASI) Summer School at the University of Colorado, Boulder. The aim of this set of lectures is neither to present in detail particle dark matter models nor to focus on the technical aspects of experiments or the (alas, necessary!) understanding of astrophysical backgrounds; rather, I try to convey and to work out order-of-magnitude estimates that can be applied to a variety of particle dark matter models and physical situations. Practice with these estimates might be perhaps appreciated by the theory-minded scholar who is not necessarily keen on the fine print of experimental setups or on the complications of astrophysical processes.

The first lecture is devoted to an introduction to the cosmology of particle dark matter, specifically in connection with the indirect detection of dark matter: what are the relevant/expected energy scales, particle products and rates? The second lecture discusses the thermal (or non-thermal) processes associated with the production of dark matter in the early universe, a few exceptions to the ``standard lore'', and the process of kinetic decoupling. The third lecture introduces indirect detection of dark matter, and details on charged cosmic rays produced by dark matter in the Galaxy. Finally, the fourth lecture discusses gamma rays and high-energy neutrinos from dark matter. Ten simple exercises are scattered through the lectures. The interested Reader might find them useful practice to master the material being discussed.

The reference list is vastly incomplete, and it primarily egotistically contains self-citations: I consider it part of a good scientific education to go the extra mile and find relevant papers on given topics of interest. This will therefore be Exercise \#11! Enjoy!

\clearpage

\section*{Lecture 1: Particle Dark Matter: zeroth-order Lessons from Cosmology\label{sec:lecture1}}
\addcontentsline{toc}{section}{Lecture 1: Particle Dark Matter: zeroth-order Lessons from Cosmology}

The fundamental (elementary) particle nature of dark matter can be probed with the detection of photons, neutrinos or charged cosmic rays that are produced by, or affected by, dark matter as an elementary particle. The key processes are:

\begin{itemize}
\item[(a)] the {\em pair annihilation} of dark matter particles (which we shall generically indicate here with the symbol $\chi$), producing Standard Model (SM) particles in the final state: $\chi+\chi\to {\rm SM}$;
\item[(b)] the {\em decay} of dark matter particles into SM particles: $\chi\to{\rm SM}$;
\item[(c)] the {\em elastic scattering} of dark matter particles off of SM particles: $\chi+{\rm SM}\to\chi+{\rm SM}$.
\end{itemize}
\noindent Other processes might exist, but are less common in the literature and in model building, and we will not entertain them here. It is important to note that none of the processes (a-c) listed above is bound to necessarily occur in any particle dark matter model: for example (b) does not occur if the dark matter is absolutely stable, and (a) and (c) can be highly suppressed, or even not occur at all, if the coupling of the dark matter sector to the SM is suppressed, or if the dark matter sector is somehow ``secluded'', or if the dark matter is not its own antiparticle, and there is no anti-dark matter around.

There exist, however, reasons to be optimistic with respect to the prospect of detecting non-gravitational signatures from dark matter: firstly, some of the best motivated (from a theoretical standpoint) extensions to the SM encompassing a dark matter candidate $\chi$ predict a coupling of $\chi$ to SM particles that would entail processes (a)-(c) or (b) at some level; secondly, there exist ``phenomenological'' reasons, chiefly the so-called WIMP miracle (to be reviewed in what follows), that imply the occurrence of some of the processes above for models where the observed abundance of dark matter is connected (via thermodynamics and cosmology) to its particle nature.

There are three key ingredients to understand indirect dark matter detection at a qualitative level, and to be able to make quantitative predictions:

\begin{enumerate}
\item {\em Production rates} of the relevant SM particles (``messengers''); this is related to the pair-annihilation or decay rate of the dark matter particle;
\item {\em Energy scale} of the SM messengers: this is set by the mass of the dark matter particle (or by its momentum, for processes of type (c) above);
\item {\em Annihilation products}: this largely model-dependent ingredient details on which SM particles are produced by the dark matter particle.
\end{enumerate} 

The rate $\Gamma_{e^\pm,\bar p, \gamma,\nu,\ldots}$ for a given SM messenger (i.e. the flux of such particle species per unit time from a unit volume $V$ containing dark matter particles) is generically the product of three factors: (1) the number of dark matter particle pairs (or of dark matter particles, for decaying $\chi$) in the volume $V$ times (2) the pair annihilation (respectively, the decay) rate, times (3) the flux of SM particles per annihilation (decay) event. In formulae:
\begin{eqnarray*}
\Gamma_{\rm SM,\ ann}&=&\left(\int\frac{\rho_{\rm DM}^2}{m_\chi^2}{\rm d} V\right)\times\left(\sigma v\right)\times\left(N_{\rm SM,\ ann}\right),\\
\Gamma_{\rm SM,\ dec}&=&\left(\int\frac{\rho_{\rm DM}}{m_\chi}{\rm d} V\right)\times\left(\frac{1}{\tau_{\rm dec}}\right)\times\left(N_{\rm SM,\ dec}\right).
\end{eqnarray*}
Interestingly for the present discussion, many key quantities ($\rho_{\rm DM},\ m_\chi,\ \sigma v$,\ldots) are potentially connected to how the dark matter was produced in the early universe. The dark matter production mechanism in the very early universe is therefore a great starting point both for model building and for eye-balling the relevant indirect detection techniques and for setting constraints.

The one quantity from cosmology which is important to have in mind is the average dark matter density in the universe, $$\bar\rho_{\rm DM}\simeq0.23\cdot \rho_{\rm crit}=\Omega_{\rm DM}\cdot \frac{3H_0^2}{8\pi G_N}.$$
It is useful in many social (as well as anti-social) situations to have on the tip of your tongue the value of this latter quantity both in ``astronomical''\footnote{Particle physicist: always a good idea to talk to astronomers; for example, I found my wife that way!} and in ``particle physics'' units: $$\rho_{\rm crit}\simeq3\times10^{10}\ \frac{M_{\odot}}{{\rm Mpc}^3}\simeq 10^{-6}\ \frac{\rm GeV}{{\rm cm}^3}.$$
From the ``astronomical'' units, we learn for example that clusters of galaxies, the largest bound dark matter structures in the universe, have typical over-densities\footnote{An ``over-density'' is a region whose average density is larger, by a certain (over-density) factor, than the overall average density.} of $10^5$, since they approximately host hundreds to thousands of galaxies, whose mass is in the $\sim 10^{12}\ M_\odot$ range; from the ``particle physics'' units, we learn that in our particular location in the Milky Way, where $\rho_{\rm DM}\sim0.3$ GeV/cm$^3$, the over-density is a factor of a few larger than in a typical cluster.

A successful framework for the origin of species in the early universe is the paradigm of thermal decoupling (see e.g. \cite{KT} and \cite{dodelson}). This framework encompasses for example the successful predictions of recombination and of Big Bang nucleosynthesis, and it describes in detail the process of cosmological neutrino decoupling. In short, thermal decoupling consists of the relevant particle interaction rate $\Gamma$, initially, at high temperatures, much larger than the Hubble expansion rate $H$, falling to a ``freeze-out'' point where $\Gamma\sim H$; after this point in time/temperature, the particle species simply ``redshifts'' its momentum and number density.

In natural units, we can think of $\Gamma=n\cdot \sigma$, with $n$ a particle number density and $\sigma$ an interaction cross section. As statistical mechanics kindly teaches us, the equilibrium number density of a particle of mass $m$ in a thermal bath of temperature $T$ has two asymptotic regimes:
\begin{eqnarray*}
n_{\rm rel}&\sim& T^3\quad{\rm for}\ \ m\ll T,\\
n_{\rm non-rel}&\sim& (mT)^{3/2}\exp\left(-\frac{m}{T}\right)\quad{\rm for}\ \ m\gg T.\\
\end{eqnarray*}
The right-hand side of $\Gamma\sim H$, i.e. $H(T)$, comes from general relativity,  and specifically from Friedmann's equation: $$H^2=\frac{8\pi G_N}{3}\rho.$$  
In the radiation dominated epoch (i.e. $T\gtrsim 1$ eV), $$\rho\simeq\rho_{\rm rad}=\frac{\pi^2}{30}\cdot g\cdot T^4,$$ with $g$ the number of relativistic degrees of freedom ($g=2$ for photons). To a decent degree of approximation, and recalling that $M_P=1/\sqrt{8\pi G_N}$, in order to eyeball when thermal decoupling occurs you can take $H\simeq T^2/M_P$. 

Let's put all of this in practice, and estimate the temperature of neutrino freeze-out. We estimate the relevant scattering cross section in the Fermi four-fermion contact interaction approximation, and we take $E\sim T_\nu$, so that $\sigma\sim G_F^2 T_\nu^2$ (where \mbox{$G_F\sim10^{-5}\ {\rm GeV}^{-2}$} is Fermi's constant). We thus have $$n\cdot \sigma=H \quad \rightarrow \quad T_\nu^3 G_F^2T_\nu^2=T_\nu^2/M_P.$$
We thus have $$T_\nu=(G_F^2 M_P)^{-1/3}\simeq(10^{-10}\times 10^{18})^{-1/3}\ {\rm GeV}\sim 1\ {\rm MeV}.$$
Among various things to be happy about, we cheerfully verify that the $T_\nu$ we found indeed satisfies $T_\nu\gg m_\nu$, which we have implicitly assumed for the form of $n(T)$, and we learn that neutrinos are {\em hot relics}, not because they are particularly attractive, but because they freeze out while they are relativistic.

Now let's calculate the relic density for a dark matter particle $\chi$. Let me introduce the notation $Y=n/s$ where $n$ is a number density and $s$ is the entropy density. In an iso-entropic universe, $s\cdot a^3={\rm constant}$, where $a$ is the universe's scale factor. $Y\sim na^3$ is thus a ``comoving'' number density. If no entropy is produced, $Y_{\rm today}= Y_{\rm freeze-out}$. In the case of hot relics, like SM neutrinos, $$Y_{\rm freeze-out}=\frac{\rho_\nu(T_\nu)}{m_\nu\cdot s(T_\nu)}$$ and $$n_{\rm today}=s_{\rm today}\times Y_{\rm freeze-out}.$$ For example, for the SM neutrinos, the fraction of the universe's critical density times $h^2$ (where $h$ is today's Hubble constant in units of 100 km/s/Mpc -- in practice $h^2\simeq 0.5$) in a SM neutrino species is $$ \Omega_\nu h^2=\frac{\rho_\nu}{\rho_{\rm crit}}h^2\simeq\frac{m_\nu}{91.5\ {\rm eV}}.$$
While the normalization depends on the relevant cross section, it is a general fact that  {\em a hot relic's thermal relic abundance scales linearly with the relic's mass}. For a weakly interacting dark matter particle, requiring that the thermal dark matter density be less or equal than the observed matter density leads to the so-called Coswik-McClelland limit \cite{coswikmcc} on the mass of a hot dark matter relic.

\begin{center}
\fbox {
    \parbox{0.8\linewidth}{
    {\bf Exercise \#1:} Calculate $T_{\rm f.o.}$ for the $p\bar p$ annihilation reaction (you can use $\sigma\sim m_\pi^{-2}$) and estimate the relic proton/antiproton density; is this a hot relic problem? Compare what you find with the observed ``baryon asymmetry''.
    }
}
\end{center}

A {\em cold relic} is one for which the freeze-out temperature is much lower than the mass of the particle, which thus decouples in the non-relativistic regime. An illustrative example of a cold relic is a ``heavy'' neutrino, with a mass $m_N\gg1$ MeV. In this case, the appropriate asymptotic form for the equilibrium number density is the non-relativistic limit $$n\sim\left(m_\chi T\right)^{3/2}\exp\left(-\frac{m_\chi}{T}\right).$$ Again, the condition $n\sigma\sim H$ yields 
\begin{equation}\label{eq:intfo}
n_{\rm f.o.}\sim\frac{T_{\rm f.o.}^2}{M_P\cdot\sigma}.
\end{equation}
 Let us call $m_\chi/T\equiv x$; when dealing with cold relics, we are thus working in the $x\gg1$ regime. We can re-cast the condition $n\cdot \sigma\sim H$ as $$\frac{m_\chi^3}{x^{3/2}}e^{-x}=\frac{m_\chi^2}{x^2\cdot M_P\cdot \sigma}.$$ We thus need to solve 
\begin{equation}\label{eq:fo}
\sqrt{x}\cdot e^{-x}=\frac{1}{m_\chi\cdot  M_P\cdot \sigma}\sim\frac{1}{10^2\cdot 10^{18}\cdot 10^{-6}}\sim 10^{-14},
\end{equation} 
where I've substituted for the nominal values of an ``electro-weak interacting'' relic, with $\sigma\sim G_F^2m_\chi^2$ and $m_\chi\sim 10^2$ GeV. Numerically, for the range $10^{-10}...10^{-20}$ for the right-hand side of Eq.~(\ref{eq:fo}), the resulting $x_{\rm f.o.}\simeq 20...50$. Now, 
\begin{equation*}
\Omega_\chi=\frac{m_\chi\cdot n_\chi(T=T_0)}{\rho_c}=\frac{m_\chi\ T_0^3}{\rho_c}\frac{n_0}{T_0^3},
\end{equation*}
with $T_0=2.75$ K $\sim10^{-4}$ eV. Since for an iso-entropic universe $aT\sim$const, $$\frac{n_0}{T_0^3}\simeq\frac{n_{\rm f.o.}}{T_{\rm f.o.}^3}$$ we have
\begin{equation*}
\Omega_\chi=\frac{m_\chi\ T_0^3}{\rho_c}\frac{n_{\rm f.o.}}{T_{\rm f.o.}^3}=\frac{T_0^3}{\rho_c}x_{\rm f.o.}\left(\frac{n_{\rm f.o.}}{T_{\rm f.o.}^2}\right)=\left(\frac{T_0^3}{\rho_c\  M_P}\right)\frac{x_{\rm f.o.}}{\sigma}.
\end{equation*}
where I used Eq.~(\ref{eq:intfo}) in the last step. The equation above can be then cast, plugging in the numbers for the various constants, as
\begin{equation}\label{eq:miracle}
\left(\frac{\Omega_\chi}{0.2}\right)\simeq\frac{x_{\rm f.o.}}{20}\left(\frac{10^{-8}\ {\rm GeV}^{-2}}{\sigma}\right),
\end{equation}
a relation that many refer to as ``miraculous''. Often, Eq.~(\ref{eq:miracle}) is quoted with the thermally-averaged product of the cross section times velocity $\langle\sigma v\rangle$ (we will understand why this is, and what a thermal average is, in the next lecture), instead of the simple cross section $\sigma$. Since $v\sim c/3$ for $x\sim20$, one has $$\langle\sigma v\rangle\sim 10^{-8}\ {\rm GeV}^{-2}\left(3\times10^{-28}\ {\rm GeV}^2\ {\rm cm}^2\right)\ 10^{10}\ \frac{\rm cm}{\rm s}=3\times10^{-26}\ \frac{{\rm cm}^3}{\rm s}.$$
\begin{center}
\fbox {
    \parbox{0.8\linewidth}{
    {\bf Exercise \#2:} Convince yourself that $v\sim c/3$ for $x\sim20$.
    }
}
\end{center}

The $\langle\sigma v\rangle\sim3\times10^{-26}\ {\rm cm}^3/ {\rm s}$ is a ``magic'' number definitely worth keeping in mind! 

Is the ``magic number'' we just found unique and peculiar to the electroweak scale? Not at all! Let us remind ourselves which ingredients we used to get the ``right'' relic density:
\begin{itemize}
\item[(i)] the condition for having a cold relic, $m_\chi\cdot \sigma\cdot M_P\gg 1$;
\item[(ii)] a cross section $\sigma\sim10^{-8}\ {\rm GeV}^{-2}$.
\end{itemize}
Now, suppose that the cross section be $$\sigma\sim\frac{g^4}{m_\chi^2},$$ with $g$ some coupling. Condition (ii) essentially enforces that $$g^2\sim\frac{m_\chi}{10\ {\rm TeV}}$$ independently of which scale $m_\chi$ is at. Now go back to condition (i): this reads $$M_P\gg\frac{1}{m_\chi\cdot \sigma}=\frac{m_\chi^2}{m_\chi\cdot (g^2)^2}\sim\frac{10^8\ {\rm GeV}^2}{m_\chi}$$ if condition (ii) holds. Therefore, condition (i) implies, with condition (ii), that $m_\chi\gg0.1$ eV. Therefore, thermal freeze-out giving the ``right'' relic abundance is not peculiar to the electroweak scale, as reiterated recently in the literature (see e.g. the ``WIMPless'' miracle of Ref.~\cite{wimpless}). However, since $$\sigma_{\rm EW}\sim G_F^2 T_{\rm f.o.}^2\sim G_F^2(\frac{E_{\rm EW}}{20})\sim10^{-8}\ {\rm GeV}^{-2},$$ the electroweak scale is quite a ``natural'' place (whatever natural means) for the miracle to occur!

Is there any {\em upper} limit to the particle dark matter mass in the cold thermal relic scheme? Indeed there is! The coupling constant $g$ cannot be arbitrarily large (a condition that can also be rephrased in terms of a unitarity limit in the partial wave expansion \cite{partialwave}; note that caveats to the unitarity argument do exist, and this limit can be evaded! (I suggest you read Ref.~\cite{partialwave} and think about how to do that)). Roughly, $$\sigma\lesssim\frac{4\pi}{m_\chi^2},$$ which implies $$\frac{\Omega_\chi}{0.2}\gtrsim10^{-8}\ {\rm GeV}^{-2}\cdot {\frac{m_\chi^2}{4\pi}}.$$ Therefore, demanding $\Omega_\chi\lesssim 0.2$ implies $$\left(\frac{m_\chi}{120\ {\rm TeV}}\right)^2\lesssim1,$$ or $m_\chi\lesssim 120$ TeV.

Is there, similarly, a {\em lower} limit in the cold thermal relic scheme? We commented above on the general limit, for arbitrarily low cross sections, $m_\chi\gg0.1$ eV. But suppose now we have in mind a particle that interacts via electroweak interactions, for example, again, a massive neutrino with $\sigma\sim G_F^2\ m_\chi^2$. In this case $$\Omega_\chi h^2\sim0.1\frac{10^{-8}}{{\rm GeV}^{-2}}\cdot\frac{1}{G_F^2\ m_\chi^2}\sim0.1\left(\frac{10\ {\rm GeV}}{m_\chi}\right)^2.$$ This implies that $m_\chi\gtrsim10$ GeV for weakly interacting massive particles (WIMPs) -- a limit known in the literature as the {\em Lee-Weinberg limit} \cite{leeweinberg}. 

\begin{figure}[t]
\centering
\begin{minipage}{0.45\textwidth}
\hspace*{-1cm}\includegraphics*[scale=0.5]{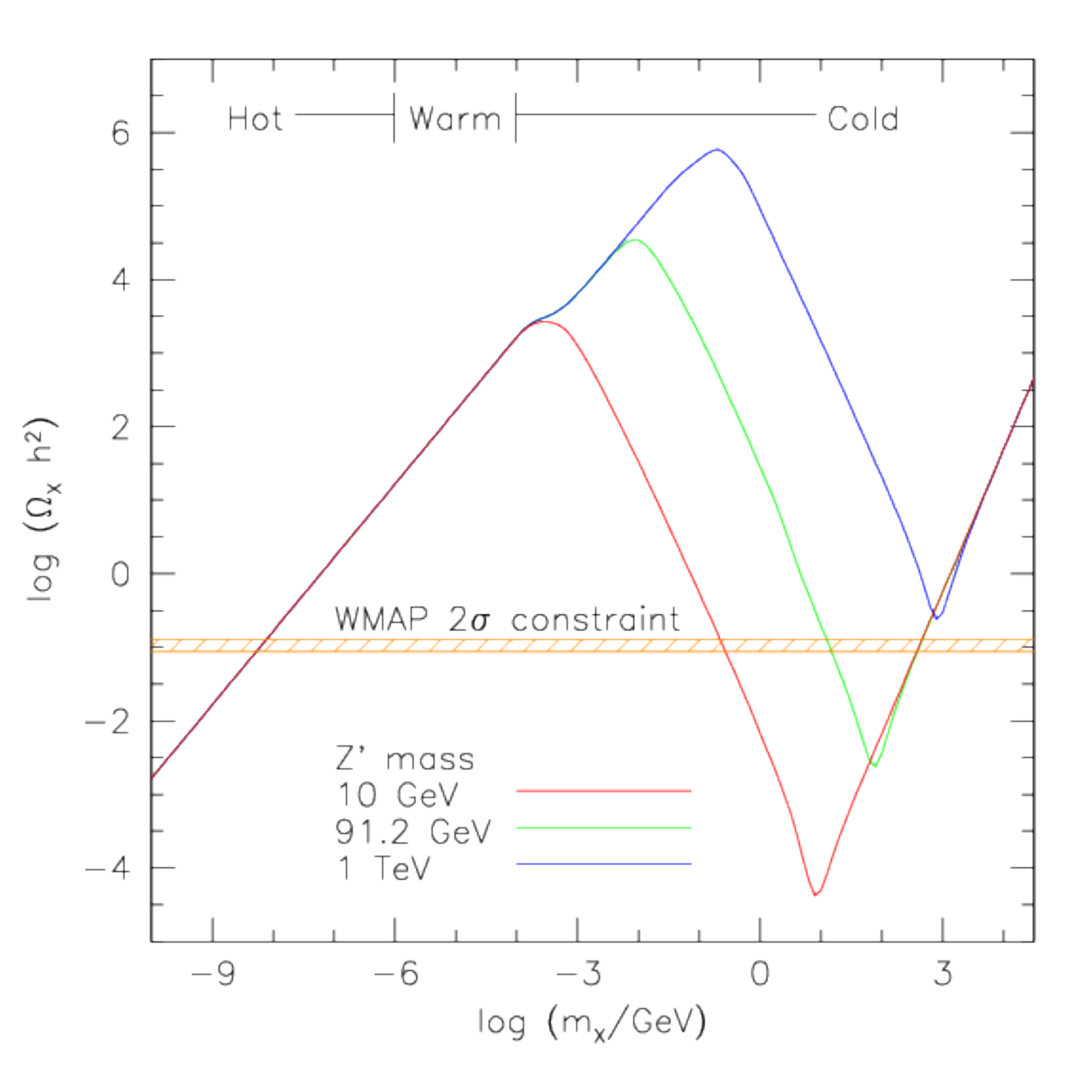}
\end{minipage}
 \hfill  
\caption{The thermal relic density of a relic that pair-annihilates with the cross section of Eq.~(\ref{eq:xs}), for three values of $m_{Z^\prime}$. From Ref.~\cite{baltzrev}.
\label{fig:relicdensity}}
\end{figure}

Fig.~\ref{fig:relicdensity}, from Ref.~\cite{baltzrev}, illustrates the thermal relic density of a weakly interacting massive particle as a function of the particle's mass. The cross section is assumed to be of the form 
\begin{equation}\label{eq:xs}
\sigma\sim\frac{m_\chi^2}{(s-m_{Z^\prime}^2)^2+m^4_{Z^\prime}},
\end{equation} 
with $s$ the total center of mass energy squared. The mass of the mediator $Z^\prime$ is taken to be 10 GeV, 91.2 GeV (the $Z$ mass) and 1 TeV. The asymptotic hot and cold relic behaviors are clearly visible and match the predictions we made above: a hot relic density scales linearly with mass, a cold relic with $m_\chi\gg m_{Z^\prime}$ has $\Omega\sim1/\sigma\sim m_\chi^2$, and a cold relic in the regime where $m_\chi\ll m_{Z^\prime}$ has $\Omega\sim m_{Z^\prime}^4/m_\chi^{2}$. 

\begin{figure}[t]
\centering
\begin{minipage}{0.45\textwidth}
\hspace*{-1cm}\includegraphics*[scale=0.5]{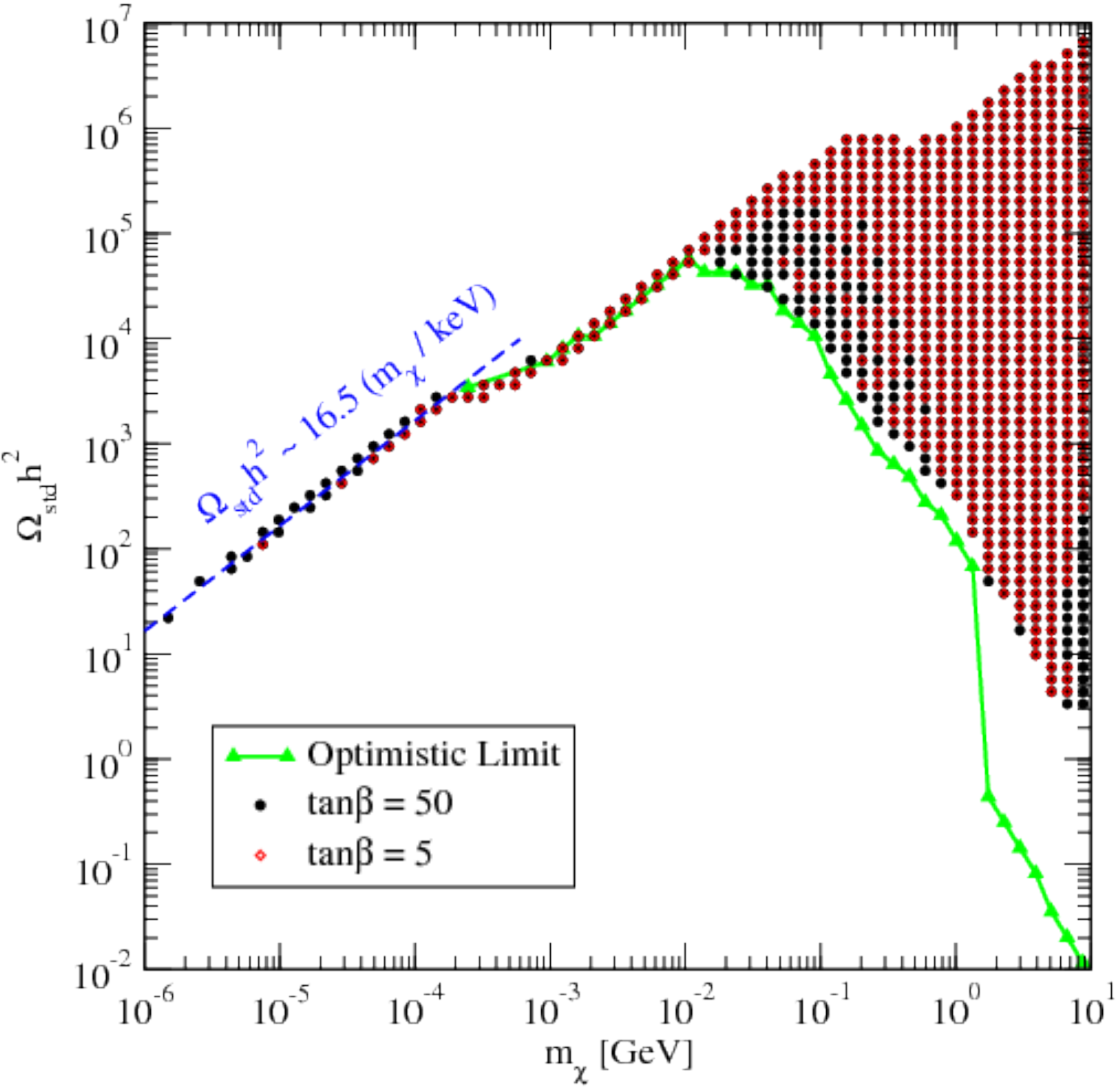}
\end{minipage}
 \hfill  
\caption{The thermal relic density of the lightest neutralino in the MSSM, as a function of the neutralino mass. From Ref.~\cite{relicsusyref}.
\label{fig:relicsusy}}
\end{figure}

In theories with a more complicated dark matter pair-annihilation cross section than what appears in Eq.~(\ref{eq:xs}), for example for the lightest neutralino in the minimal supersymmetric extension of the Standard Model (MSSM), various effects blur the simple connection between mass and cross section/relic density, resulting in a much larger spread of results. The general feature of a lower limit of about 1-10 GeV is, however, rather resilient, as shown in Fig.~\ref{fig:relicsusy}, from Ref.~\cite{relicsusyref}, where I scanned generously over the relevant MSSM parameter space.  The black dots indicate points with $\tan\beta$ (the ratio of the vacuum expectation values of the two Higgses in the MSSM) fixed to 50, while the red points have $\tan\beta=5$. I invite the SUSY-afecionado to read my paper for details of the scan. The $x$-axis indicates the lightest neutralino mass, while the $y$-axis is the thermal relic density calculated in a standard cosmology without entrooy injection or a modified Hubble expansion rate. I stretched the relevant parameters as hard as I could to obtain the ``optimistic limit'' green line. That line shows that for $\Omega h^2\simeq0.1$ the lightest neutralinos I find are in the few GeV range.

Let us now ask the general question: how light can a WIMP be? So far, we assumed an iso-entropic universe. Suppose at some point after a WIMP has frozen out and is thus decoupled from the universe's thermal bath, the entropy density changes from $s\to\gamma\cdot s$, $\gamma>1$ from e.g. decaying relics (such as relic gravitinos, moduli, \ldots) or from a first order phase transition (see e.g. Ref.~\cite{maxphase}). Then $$Y_{\rm today}\to\frac{Y_{
\rm today}}{\gamma}\quad {\rm and}\quad \Omega_\chi\to\frac{\Omega_\chi}{\gamma}.$$
For a sufficiently large $\gamma$, the relic abundance of almost any over-abundant relic WIMP can be ``diluted'' enough to match the observed dark matter density. For example, in supersymmetry the lightest neutralino can be almost arbitrarily light as long as it is bino-like (to prevent an excessively light associated chargino) and if sfermions are sufficiently heavy to suppress energy loss mechanisms in stars (for example $e^+ e^-\to\chi\chi$). The additional requirement is of course that the entropy injection happen at a temperature smaller than the neutralino freeze-out, which sets a (weak) constraint on the neutralino mass \cite{relicsusyref}: in order to maintain the successful predictions of light elemental abundances, entropy injection cannot happen too close to the era of Big Bang Nucleosynthesis.

\clearpage

\section*{Lecture 2: WIMP Relic Density, a Closer Look \label{sec:lecture2}}
\addcontentsline{toc}{section}{Lecture 2: WIMP Relic Density, a Closer Look}

The dark matter literature is flooded with the symbol $\langle\sigma  v\rangle$, short for the {\em zero-temperature thermally-averaged pair-annihilation cross section times velocity} -- but do we really understand what this symbol indicates, and where it comes from? How is the ``thermal average'' defined? What is $v$? is it a relative velocity? but a relative velocity is not Lorentz-invariant! etc...

The starting point for a closer look at the WIMP relic density is the Boltzmann equation, that can be symbolically cast as \cite{KT}: 
\begin{equation}\label{eq:boltz}
\hat L[f]=\hat C[f],
\end{equation}
where $f=f(\vec p,\vec x, t)$ is the phase space density, $\hat L$ is the {\em Liouville operator} describing the change in time of the phase space density, and $\hat C$ is the {\em collision operator} describing the number of particles per phase-space volume lost or gained per unit time. For those (like me) who need a refresher on the Liouville operator, its non-relativistic form reads $$\hat L_{\rm NR}=\frac{\rm d}{{\rm d}t}+\frac{{\rm d}\vec x}{{\rm d}t}\vec\nabla_x+\frac{{\rm d}\vec v}{{\rm d}t}\vec\nabla_v,$$
while in its covariant form it reads
$$\hat L_{\rm cov}=p^\alpha\frac{\partial}{\partial x^\alpha}-\Gamma^\alpha_{\beta\gamma}\ p^\beta\  p^\gamma\frac{\partial}{\partial p^\alpha}.$$
In a homogeneous and isotropic cosmology (also known in the trade's slang as a Friedman-Robertson-Walker universe), $$f(\vec x, \vec p,t)\to f(|\vec p|,t)\quad {\rm or,\ equivalently:}\quad f(E,t).$$
Also, $\hat L$ simplifies to $$\hat L[f]=E\frac{\partial f}{\partial t}-\frac{\dot a}{a}|\vec p|^2\ \frac{\partial f}{\partial E}.$$
We are interested in particle number densities, defined by $$n(t)=\sum_{\rm spin}\int\frac{{\rm d}^3p}{(2\pi)^3}f(E,t).$$
We will thus take Eq.~(\ref{eq:boltz}) and consider (calling $g$ the number of spin degrees of freedom) $$\int L[f]\cdot g\frac
{{\rm d}^3p}{(2\pi)^3}=\frac{{\rm d}n}{{\rm d}t}+3H\cdot n$$ where we introduced $H=\dot a/a$ and integrated by parts using $$\frac{1}{a^3}\frac{{\rm d}}{{\rm d}t}\left(a^3\cdot n\right)=\frac{{\rm d}n}{{\rm d}t}+3H\cdot n.$$ Cleaning up the right-hand side of the Boltzmann equation is a bit messier and I recommend the classic paper by Gondolo and Gelmini, Ref.~\cite{gondologelmini}. For definiteness, let us consider a process of the type $1+2\to3+4$ where we are interested in the number density of species 1, and where we assume that species 3 and 4 are in thermal equilibrium. The right-hand side of Eq.~(\ref{eq:boltz}) can then be cast as:
\begin{equation*}
g_1\int \hat C[f_1]\frac{{\rm d}^3p}{(2\pi)^3}=-\langle\sigma\cdot v_{\rm M\o l}\rangle\left(n_1n_2-n_1^{\rm eq}n_2^{\rm eq}\right),
\end{equation*}
where $n_{1,2}$ are the number densities, while $n_{1,2}^{\rm eq}$ indicate the equilibrium number densities, and where $$\sigma=\sum_f\sigma_{12\to f}$$ indicates the invariant, unpolarized total cross section for processes $1+2\to$ any final state $f$ in thermal equilibrium, and where, finally, the ``M\o ller velocity\footnote{Usually the M\o ller velocity is defined as $v_{\rm M\o l}=\left((\vec v_1-\vec v_2)^2-(\vec v_1\times \vec v_2)^2\right)^{\frac{1}{2}}$.}'' is defined in the following covariant form:
\begin{equation*}
v_{\rm M\o l}\equiv\frac{\sqrt{(p_1\cdot p_2)^2-m_1^2m_2^2}}{E_1\ E_2}.
\end{equation*}
A couple of comments:

\begin{enumerate}

\item Notice that $v_{\rm M\o l} \ n_1\ n_2$ is a Lorentz invariant quantity;

\item Notice that in the rest frame of 1 (or 2; what we can think of as the ``lab frame''), $v_{\rm M\o l}\to v_{\rm rel}=|\vec v_1-\vec v_2|$, where e.g. $\vec v_1=\vec p_1/E_1$ etc.
\end{enumerate}

Last ingredient: the thermal average: this is defined by the expression:

\begin{equation}\label{eq:thave}
\langle\sigma\cdot v_{\rm M\o l}\rangle=\frac{\int \sigma\cdot v_{\rm M\o l}\  e^{-E_1/T}e^{-E_2/T}\ {\rm d}^3p_1\ {\rm d}^3p_2}{\int e^{-E_1/T}e^{-E_2/T}\ {\rm d}^3p_1\ {\rm d}^3p_2}.
\end{equation}

\begin{center}
\fbox {
    \parbox{0.8\linewidth}{
    {\bf Exercise \#3:} Evaluate the denominator of Eq.~(\ref{eq:thave}) for $m_1=m_2$.
    }
}
\end{center}
The diligent Reader who carried out the exercise proposed above found that the denominator of Eq.~(\ref{eq:thave}), for $m_1=m_2=m$ (the pair-annihilation case relevant for us) reads $${\int e^{-E_1/T}e^{-E_2/T}\ {\rm d}^3p_1\ {\rm d}^3p_2}=\left(4\pi m^2 T K_2\left(\frac{m}{T}\right)\right)^2,$$ where $K_2$ is the modified Bessel function of the second order. The numerator reads, instead,
\begin{equation}\label{eq:xsint}
\int \sigma\cdot v_{\rm M\o l} e^{-E_1/T}e^{-E_2/T}\ {\rm d}^3p_1\ {\rm d}^3p_2=\int_{4m^2}^\infty\ \sigma(s-4m^2)\sqrt{s}K_1\left(\frac{\sqrt{s}}{T}\right){\rm d}s,
\end{equation}
a ``convolution'' of the cross section evaluated at $s-4m^2$, where $s$ is the center-of-mass total energy, and a temperature-dependent thermal kernel. This form is key to understand important caveats to, e.g. the magic relation that implies that $\langle \sigma v\rangle=3\times 10^{-26}\ {\rm cm}^3/$s gives the correct thermal relic density. From now on, I will suppress the subscript M\o l and intend always that $v\to v_{\rm M\o l}$.

\subsection*{Caveats to the Standard Story}

A classic paper that discusses departures from the vanilla calculation of the WIMP relic density described in the previous lecture is Ref.~\cite{gs} by Griest and Seckel, ``Three exceptions in the calculation of relic abundances''. The three memorable exceptions are:
\begin{enumerate}
\item Resonances;
\item Thresholds;
\item Co-annihilations.
\end{enumerate}
Resonant annihilation through a particle with the right quantum numbers and a mass $m_A\simeq 2m_\chi$, found for example in the so-called ``funnel'' region of the (soon to be gone, thanks LHC!) minimal supergravity/constrained MSSM model, can be relevant either if $m_\chi\gtrsim m_A/2$ or if $m_\chi\lesssim m_A/2$. In the first case, the cross section peaks at $s=m_A^2$ and is thus most relevant at temperatures $T_{\rm res}\simeq m_A^2/(6m_\chi)$. 
\begin{center}
\fbox {
    \parbox{0.8\linewidth}{
    {\bf Exercise \#4:} Show that $\langle s\rangle\simeq 4m_\chi^2+6m_\chi T$.
    }
}
\end{center}
If $T_{\rm res}\simeq T_{\rm f.o.}$ the resonance is extremely important at freeze-out, and hence for the thermal relic density; the pair annihilation cross section today, the one that we care about for indirect dark matter detection rates, will however be potentially much lower! In this case, the freeze-out cross section might be close to $\langle \sigma v\rangle=3\times 10^{-26}\ {\rm cm}^3/$s, but (potentially) the $T=0$ cross section $\langle \sigma v\rangle_0\ll3\times 10^{-26}\ {\rm cm}^3/$s.

If $m_\chi\lesssim m_A/2$, in the integral of Eq.~(\ref{eq:xsint}) the cross section is always maximal for $T=0$, the resonance can be (or not) subdominant at freeze-out, but we are in the (lucky if one wants a signal, unlucky if one wants to hide it!) circumstance where the $T=0$ cross section$\langle \sigma v\rangle_0\gg3\times 10^{-26}\ {\rm cm}^3/$s. It must be noted that this discussion has model-dependent caveats: for example, in supersymmetry the lightest neutralinos are Majorana particles, and in a purely $s$-wave annihilation (at $T=0$) a pair of neutralinos are in a $CP$-odd state. Therefore, for example, the pair annihilation via a $CP$-even particle (such as a neutral $CP$-even Higgs) cannot contribute to the $T=0$ pair annihilation cross section!

Thresholds affect the relation between the freeze-out and $T=0$ pair-annihilation cross sections in an obvious way: the cross section in Eq.~(\ref{eq:xsint}) suddenly increases as, e.g. $s>4m_t^2$, where $m_t$ is the particle in which pair our annihilating particle can go, $\chi\chi\to \bar t t$ (think e.g. of $m_\chi\lesssim m_t$ where $t$ is the Standard Model top quark). Thresholds therefore always imply $\langle \sigma v\rangle_{\rm f.o.}>\langle \sigma v\rangle_0$.

Co-annihilation occur for particles whose freeze-out process is tangled with that of other particle species with a close enough mass so that the two freeze-out episodes are inter-connected. A necessary condition is that this second co-annihilating species 2 have a mass such that at freeze-out the  Boltzmann suppression of its equilibrium number density is not dramatic. In formulae, we want $m_2-m_1\lesssim T_{\rm f.o.}$. In this case the relevant cross section is an ``effective'' cross section that includes the appropriately Boltzmann-weighed contribution from ($N$) co-annihilating particles, i.e., with obvious notation (if this is not obvious see the extensive and clear discussion of Ref.~\cite{gondoloedsjo})
\begin{equation*}
\langle \sigma v\rangle \ \to \ \langle \sigma_{\rm eff} v\rangle=\frac{\sum_{i,j=1}^N\sigma_{ij}\exp\left(-\frac{\Delta m_i+\Delta m_j}{T}\right)}{\sum_{i=1}^N g_i \exp\left(-\frac{\Delta m_i}{T}\right)}.
\end{equation*}
In the equation above, $\Delta m_i$ indicates the difference in mass between particle $i$ and the lightest particle (to which $i$, eventually, decays). Note that the denominator counts the effect of the additional degrees of freedom, suitably weighed. 

Coannihilation comes in two varieties, that I like to call ``{\em parasitic}'' and ``{\em symbiotic}''. If the additional degrees of freedom annihilate ``less efficiently'' than the particle whose number density we are interested in, then the coannihilating particles will have a parasitic effect, and produce a smaller effective pair-annihilation cross section. This is the typical case with, for example, the lightest Kaluza-Klein excitation of universal extra dimensions (UED; for a review see my own review! Ref.\cite{uedreview}). UED has a very compressed spectrum of particles (Kaluza-Klein modes of Standard Model particles, whose mass differs from the compactification scale by loop corrections or corrections of the order of the Standard Model particle mass) above the stable dark matter particle candidate. This large collection of particles brings many additional effective degrees of freedom which outweigh the corresponding additional contribution to the pair-annihilation cross section, rendering UED a prototypical example of parasitic coannihilation.

An example of ``symbiotic'' co-annihilation is provided by a nimble particle that annihilates efficiently and that doesn't carry a large number of degrees of freedom. An example is co-annihilation of the lightest neutralino of the MSSM with the scalar partner of the tau lepton, or ``stau'' (but other equally good examples are co-annihilation with charginos or stops or other sfermions). Unfortunately, collider searches indicate that the days of the ``stau coannihilation region'' may be counted \cite{staudeath}.

So far we have dealt with exceptions to the left-hand side of the Boltzmann equation (or of its short-hand version $\Gamma=n\cdot \sigma = H$). What happens if we fiddle around with the right-hand side instead, i.e. with $H$, the expansion history of the universe? To avoid self-promoting my own papers again, I will point you to the following example: a cosmology with a ``quintessence'' field that provides a dynamical dark energy term, whose impact on the relic density of WIMPs was first studied by Salati in Ref.~\cite{salati}\footnote{See also Ref.~\cite{myquint}. Couldn't resist.}.

Let $\phi$ be the quintessence field, a spatially homogeneous real, scalar field. The field energy density and pressure are
\begin{eqnarray}
\rho_\phi&=&\frac{1}{2}\left(\frac{{\rm d}\phi}{{\rm d}t}\right)^2+V(\phi)\\
P_\phi&=&\frac{1}{2}\left(\frac{{\rm d}\phi}{{\rm d}t}\right)^2-V(\phi)
\end{eqnarray}
An example of a suitable potential that exhibits the desired ``tracking'' behavior (for appropriate initial conditions), i.e. whose energy density tracks dynamically the dominant energy density component, is $$V(\phi)=M_P^4\exp\left(-\frac{\lambda \phi}{M_P}\right).$$
The field's equation of state $w=P_\phi/\rho_\phi$ moves from $w=-1$ in the ``kination'' phase, where the kinetic energy term dominates, to $w=+1$ in the ``cosmological constant'' phase, where $V$ dominates. Tracking helps explain the coincidence problem $\Omega_\Lambda=\Omega_\phi\sim\Omega_M$, although fine-tuning is not eliminated (as it creeps back in via the field's initial conditions). 

Noticing that $\rho_\phi\sim a^{-3(1+w)}$, in the kination phase $\rho\sim a^{-6}$ and therefore the universe is kination-dominated as sufficiently early times, with $$H\sim\frac{T^2}{M_P}\frac{T}{T_{\rm KRE}},$$ where $T_{\rm KRE}$ stands for the temperature of kination-radiation equality. To be relevant for the relic density of a particle species decoupling at $T=T_{\rm f.o.}$, kination must dominate before and at freeze-out, hence $T_{\rm KRE}>T_{\rm f.o.}$. However, to avoid disrupting Big Bang Nucleosynthesis, we must also require that $T_{\rm KRE}<T_{\rm BBN}\sim 1$ MeV.

In a kination-dominated universe, freeze-out works, schematically, exactly as we described in the previous section, and
\begin{equation*}
\Omega_\chi^{\rm quint}=\frac{T_0^3}{M_P\cdot \rho_c}x_{\rm f.o.}\left(\frac{n_{\rm f.o.}}{T_{\rm f.o.}^2}\right),
\end{equation*}
but now the freeze-out condition reads
$$n_{\rm f.o.}\langle\sigma\ v\rangle\sim\frac{T^2}{M_P}\frac{T}{T_{\rm KRE}}.$$
We therefore have that 
$$ \frac{n_{\rm f.o.}}{T_{\rm f.o.}^2}\sim\frac{1}{M_P\ \langle\sigma\ v\rangle}\frac{T_{\rm f.o.}}{T_{\rm KRE}}.$$
To first order, the enhancement factor of the thermal relic density in the presence of quintessence above the standard thermal relic density is thus
$$\frac{\Omega_\chi^{\rm quint}}{\Omega_\chi^{\rm standard}}\sim\frac{T_{\rm f.o.}}{T_{\rm KRE}}\lesssim\frac{m_\chi}{20}\frac{1}{T_{\rm BBN}}\sim 10^4\frac{m_\chi}{100\ {\rm GeV}}.$$
With more accurate calculations the enhancement factor is found to be potentially as large as $10^6$ \cite{salati, myquint}.

The dark matter production mechanism can naturally be non-thermal, or it can arise from an asymmetry (as is probably the case for the origin of baryonic matter). Suppose a particle species $\psi$, with $m_\psi>m_\chi$ is produced in the early universe with an abundance $\Omega_\psi$, and that $\psi$ decays to $\chi$, which is the stable dark matter particle, at a temperature when $\chi$ is out of equilibrium. The relic density that $\chi$ will then inherit (up to contribution from the decay of other particle species and from thermal production etc.) is simply $$\Omega_\chi\simeq\Omega_\psi\frac{m_\chi}{m_\psi},$$ where the $\simeq$ sign indicates that additional effects (such as some entropy production in the decay process) can enter.

As you calculated in Ex.~\#1, the thermal abundance of protons and antiprotons is almost ten orders of magnitude smaller than the observed abundance. If an asymmetry is present, then the observed proton density can be inherited entirely from the asymmetry itself. The same could hold for the dark matter sector. Many have envisioned the possibility of explaining the coincidence $\Omega_{\rm DM}\simeq 5\Omega_{\rm B}$ (where DM is dark matter and B is baryonic matter) by postulating that perhaps $n_{\rm DM}\sim n_{\rm B}$ and that $m_{\rm DM}\sim 5$ GeV, with some mechanism that couples an asymmetry in one sector to the other sector, or many variants on this theme. 

It is interesting to ask the question of whether any indirect dark matter detection signal could exist if dark matter originates from an asymmetry. In principle, if no anti-dark-matter is generated (and if primordial pair-annihilation got rid of all of it), then it's very hard to get any indirect signals; however, if dark matter and anti-dark matter oscillate into each other (for example because of a $\Delta N_\chi=2$ operator, such as a mass term, say $m_M\chi\chi$ ) then oscillations can populate the anti-dark matter content of the universe and residual annihilation occur \cite{buckprofumo}. Unless a symmetry in the theory explicitly prohibits such operators, oscillations generically occur, since after all we only need $$\tau_{\rm universe}\sim10^{17}\ {\rm s}\lesssim 10^{17}\ {\rm s}\ \left(\frac{10^{-41}\ {\rm GeV}}{m_M}\right),\quad {\rm i.e.}\quad m_M\gtrsim 10^{-41}\ {\rm GeV}.$$

Of course many other interesting and witty dark matter production mechanisms exist: I recommend you explore at least the classic ones that pertain to sterile neutrinos and to axions (see e.g. Ref.~\cite{KT}).

\subsection*{Kinetic Decoupling}

The dark matter thermal history in the very early universe is important not only for the calculation of the particle's relic density, but potentially also for the formation of matter structure in the universe, especially for (cold) WIMPs. In the early universe, elastic scattering processes such as \mbox{$\chi f\leftrightarrow\chi f$}, where $f$ is a Standard Model fermion, keep the dark matter particle in kinetic equilibrium even {\em after} chemical decoupling (i.e. when $\Gamma_{\chi\chi\leftrightarrow X}\ll H$). The reason is that the target densities for the processes that keep the dark matter in chemical versus kinetic equilibrium are vastly different after chemical decoupling:
\begin{eqnarray*}
\chi\chi\leftrightarrow ff\quad&\to& \Gamma=n_{\rm non-rel}\cdot \sigma\\ 
\chi f\leftrightarrow \chi f\quad&\to& \Gamma=n_{\rm rel}\cdot\sigma
\end{eqnarray*}
with $n_{\rm non-rel}$ exponentially suppressed! Let us now estimate the kinetic decoupling temperature of a WIMP. We shall assume a default electroweak cross section
$$\sigma_{\chi f\leftrightarrow \chi f}\sim G_F^2 T^2.$$
We have to account for the fact that the WIMP is non-relativistic after chemical decoupling and that momentum transfer between the WIMP and the thermal bath becomes ``inefficient'', in a sense to be made quantitative with a simple estimate. The typical momentum transfer per collision is $\delta p\sim T$, while the WIMP momentum in the non-relativistic regime satisfies the relation $$\frac{p^2}{2m}\sim T$$ so that $p\sim\sqrt{m_\chi T}$. Momentum transfer is a stochastic process, so it takes $$N=\left(\frac{\delta p}{p}\right)^2\sim\frac{T^2}{m_\chi T}=\frac{T}{m_\chi}$$ collisions to establish kinetic equilibrium.

To calculate kinetic decoupling, we thus want to compare
$$n_{\rm rel}\cdot\sigma_{\chi f\leftrightarrow \chi f}\left(\frac{\delta p}{p}\right)^2\sim T^3\cdot G_F^2 T^2\cdot\frac{T}{m_\chi}\sim H\sim \frac{T^2}{M_P}.$$
We thus find
$$T_{\rm k.d.}\sim\left(\frac{m_\chi}{M_P\cdot G_F^2}\right)^{1/4}\sim30\ {\rm MeV}\ \left(\frac{m_\chi}{100\ {\rm GeV}}\right)^{1/4}.$$
What does this imply for structure formation? Roughly, the cutoff scale of the matter power spectrum will correspond to the size of the horizon at kinetic decoupling, so 
$$
M_{\rm cutoff}\sim\frac{4\pi}{3}\left(\frac{1}{H(T_{\rm kd})}\right)^3\rho_{\rm DM}(T_{\rm kd})\sim30\ M_\oplus\left(\frac{10\ {\rm MeV}}{T_{\rm kd}}\right)^3.
$$
More precisely, the cutoff scale is set by the largest of the free-streaming versus acoustic damping scale, see e.g. \cite{bringmann} for a nice review. For typical WIMPs, we thus find that these ``protohalos'' (which correspond to the first structures that gravitationally collapse in the early universe) have a mass comparable to the Earth mass (i.e. roughly $10^{-6}\ M_\odot$). However, in specific theories the range of variations can be very significant \cite{kamionsmallscale}. This fact has potentially important consequences for indirect detection, as it feeds in the problem of calculating the {\em boost factor} (i.e. the additional contribution to the count of pairs in a halo on top of the smooth halo component) from substructure, and for the dark matter ``small-scale'' problem. 

\begin{figure}[t]
\centering
\begin{minipage}{0.45\textwidth}
\mbox{\hspace*{-5cm}\includegraphics*[scale=0.37]{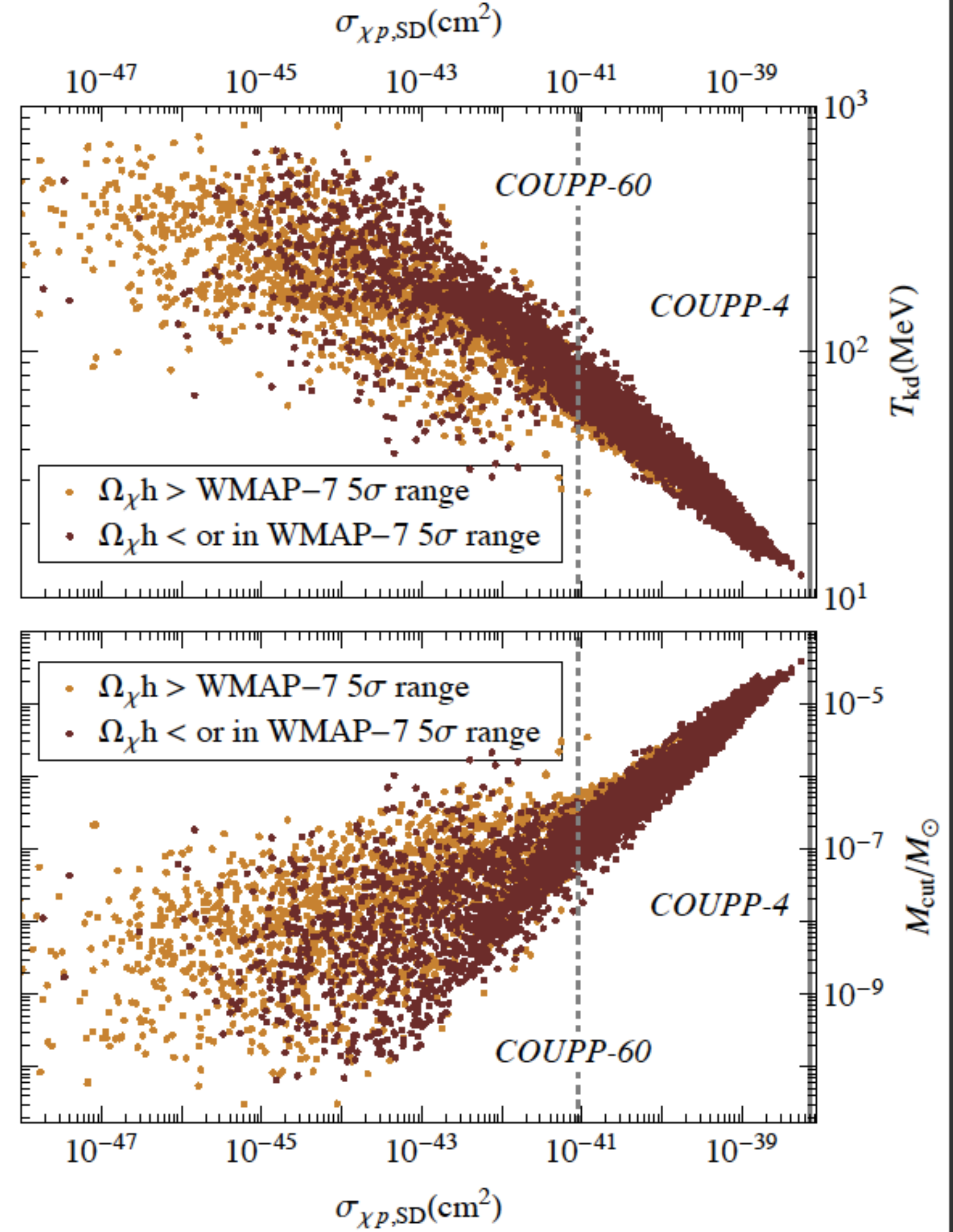}\qquad\includegraphics*[scale=0.6]{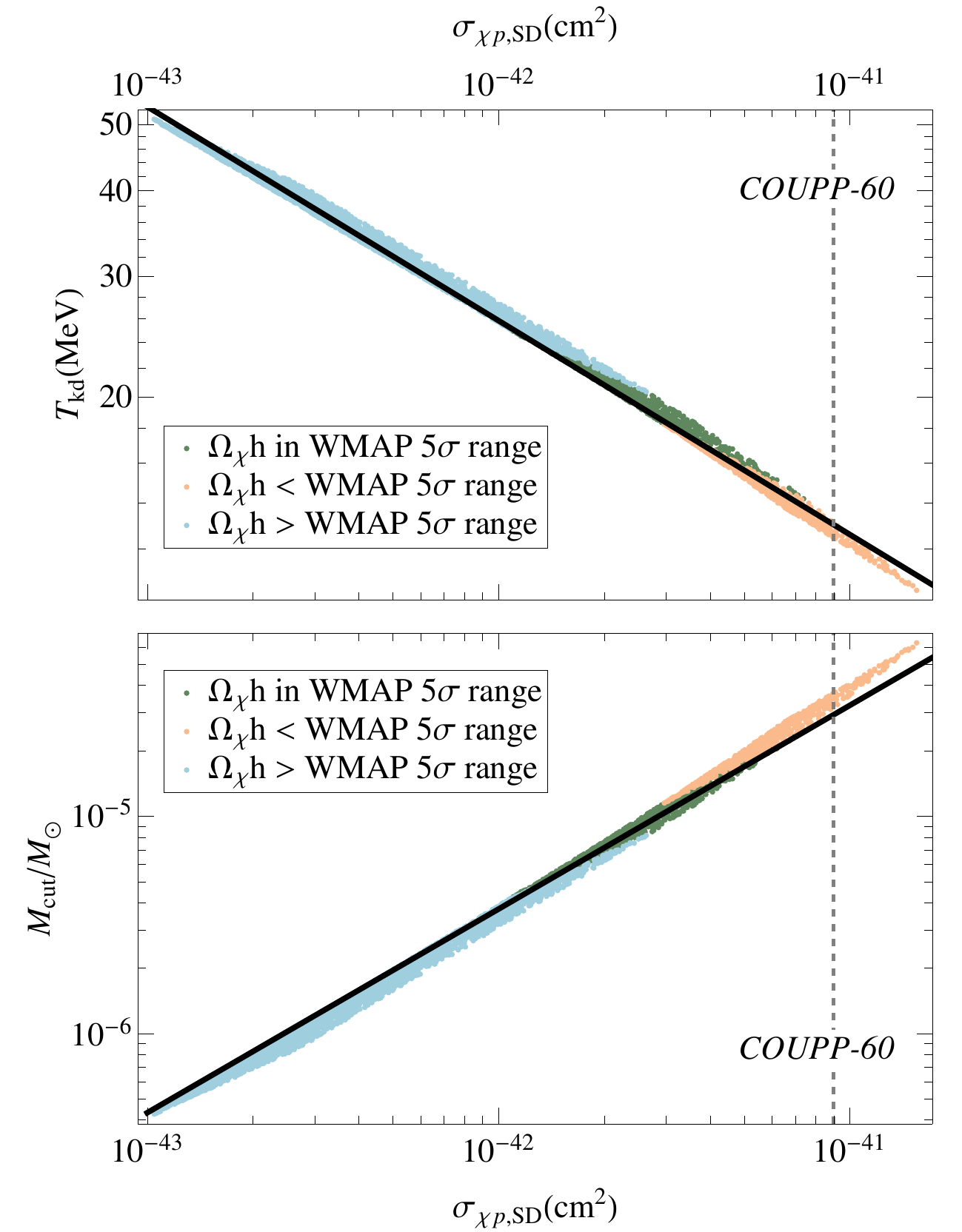}}
\end{minipage}
 \hfill  
\caption{The correlation between the spin-dependent dark matter-proton cross section and the kinetic decoupling temperature (upper panels) and the small scale cutoff mass (lower panels) in the MSSM (left panels) and in UED (right panel). From Ref.~\cite{cornell}.
\label{fig:cornell}}
\end{figure}

Is there any way to probe the size of the dark matter small-scale cutoff? In Ref.~\cite{cornell} we pointed out that if $f=q$ (i.e. the particles the dark matter scatters off of are quarks), the processes relevant for kinetic decoupling are exactly the same as those participating in dark matter direct detection. Therefore, in principle, one could correlate $M_{\rm cutoff}$ with $\sigma_{\rm direct\ det}$. There is, however, an important caveat: usually, $T_{\rm kd}<T_{\rm QCD}$, the latter symbol indicating the temperature corresponding to the QCD confinement phase transition. After confinement, the number density of hadrons is negligible compared to light leptons, and the latter dominate and control kinetic decoupling. If, however, one has a dark matter theory where ``quark-lepton universality'' holds, then a correlation is expected, and indeed is found -- both for UED and for supersymmetric dark matter \cite{cornell} as illustrated in fig.~\ref{fig:cornell}.

\clearpage

\section*{Lecture 3: Indirect Dark Matter Detection \label{sec:lecture3}}
\addcontentsline{toc}{section}{Lecture 3: Indirect Dark Matter Detection}

What have we learned thus far about annihilation processes that we could use to detect non-gravitational signals from particle dark matter?
\begin{itemize}
\item[-] $\langle\sigma v\rangle\sim3\times 10^{-26}\ {\rm cm}^3/{\rm s}$ is a good ``magic'' benchmark;
\item[-] the magic number above is not WIMP-specific, and is independent of mass, to first order;
\item[-] there exist numerous caveats, both from the particle physics side (coannihilation, resonances, thresholds,...) and from the cosmology side (quintessence, non-thermal production, asymmetry,...) to that magic number.
\end{itemize}

What about dark matter decay? If dark matter is unstable with a lifetime well in excess of the age of the universe, the decay products would also be a great way to detect non-gravitational signals from dark matter! From a theoretical standpoint, a GUT-scale or Planck-scale dark matter number-violating operators should be generic. For example for a dimension-5 operator, $$\Gamma_5\sim\frac{1}{M^2}m_\chi^3$$  the resulting lifetime 
\begin{equation}\label{eq:dim5}
\tau_5\sim 1\ {\rm s}\ \left(\frac{1\ {\rm TeV}}{m_\chi}\right)^3\left(\frac{M}{10^{16}\ {\rm GeV}}\right)^2
\end{equation}
would potentially result in an impact on BBN, but it would imply an excessively short-lived dark matter candidate\footnote{As a supplementary exercise, make sure you refresh your memory about ``natural units'' and how to convert energies into times, and convince yourself of the overall factor of 1 s in Eq.~(\ref{eq:dim5})}.

For a dimension-6 operator things look more interesting, $$\Gamma_6\sim\frac{1}{M^4}m_\chi^5$$ with a lifetime $$\tau_6\sim 10^{27}\ {\rm s}\ \left(\frac{1\ {\rm TeV}}{m_\chi}\right)^5\left(\frac{M}{10^{16}\ {\rm GeV}}\right)^4,$$ which turns out to be a very interesting lifetime range to explain the Pamela positron excess and for searches for dark matter with gamma rays, as we shall see later on.

For definiteness, let me however concentrate on dark matter annihilation. Let us now try to corner the key ingredients to make predictions for indirect searches for dark matter. First and foremost: what do we know about the dark matter particle mass, which sets the energy scale for the particles produced in an annihilation event? We saw earlier that for WIMPs a useful lower limit is provided by the Lee-Weinberg bound at about 10 GeV, while unitarity constrains WIMPs, on the large mass end, to be lighter than a few 100 TeV. I think this is a reasonable range to keep in mind, if one is wed to the notion of a weakly interacting dark matter particle.

There exist, however, a number of theoretical prejudices that have populated this field for a long time. A historically interesting one has it that WIMPs must be heavier than 40 GeV. This prejudice somehow even managed to distort how certain dark matter search experiments were optimized! It is worthwhile then to see where this prejudice comes from.

The first tenet of the ``WIMPs must be heavier than 40 GeV'' prejudice is that WIMPs are supersymmetric neutralinos. Browsing papers on dark matter (especially on {\em astro-ph}) the confusion between WIMPs and neutralinos is not unheard of. The second tenet is that there exists one universal soft supersymmetry breaking mass for all three gauginos at the GUT scale. If $M_1(M_{\rm GUT})=M_2(M_{\rm GUT})$, where $M_1$ is the soft supersymmetry breaking scale associated with the U(1)$_Y$ gaugino and $M_2$ that of the SU(2) gaugino, renormalization group evolution (and the assumption of a ``desert'' between the electroweak and the GUT scale) implies that \mbox{$M_1(M_{\rm EW})=\simeq0.4\times M_2(M_{\rm EW})$}. Now, LEP2 constrains the chargino mass to be above about half its center of mass energy, or $m_{\tilde\chi^\pm_1}\gtrsim100$ GeV. But one of the charginos (the wino-like, in SUSY slang) has a mass very close to $M_2(M_{\rm EW})$, therefore implying that the lightest, bino-like neutralino \mbox{$m_{\tilde\chi^0_1}\simeq M_1(M_{\rm EW})\gtrsim 0.4\times 100 \ {\rm GeV}=40\ {\rm GeV}$}. Amazing. Of course, GUT-scale universality, renormalization group evolution etc. are all model-dependent ingredient, not to mention the assumption that the dark matter is a neutralino...

What do we know about the annihilation final state? {\em Presque rien}, almost nothing. If the dark matter particle is a Majorana fermion, then the pair annihilation into a fermion-antifermion final state is ``$p$-wave suppressed'': $\chi\chi\to f\bar f$ requires a helicity flip, and thus the matrix element squared is proportional to the square of the fermion mass, $|M|^2\propto m_f^2$ (in just the same way as for charged pion decay -- a Majorana pair in a $l=0$ wave is in a  $CP$-odd state). As a result, pair-annihilation of Majorana dark matter into light fermions is highly suppressed. If $m_\chi<m_{\rm top}$ and if the annihilation channel $\chi\chi\to$ bosons (such as $W^+W^-$ or $hh$) is suppressed, the dominant annihilation final states are $b\bar b$ and $\tau^+\tau^-$. This explains the otherwise surprising popularity of these two final states in the literature on dark matter indirect detection\footnote{If you are a graduate student at UCSC who happened to work on dark matter, and I sit on your thesis committee, I will ask you the reason why $b\bar b$ is such a popular final state during your thesis defense. Be ready.}. Note that besides $m_b\gg m_\tau$, the bottom quark final state wins by an additional factor 3 from color. There exist, however, circumstances where the $\tau\tau$ final state can be boosted, for example with a light scalar tau in supersymmetry (in the so-called stau coannihilation region).

In UED the situation is entirely different: the particle that is usually the stable lightest Kaluza-Klein excitation is the $n=1$ mode of the hyper-charge gauge boson, or $B^{(1)}$. The matrix element squared for pair annihilation into a fermion-antifermion pair is proportional to the fourth power of the fermion's hyper-charge, $|M|^2\propto|Y_f|^4$, thus up-type quarks ($Y_{u_L}=4/3$) and charged leptons ($Y_{e_R}=2$) are the preferred annihilation modes.

If the dark matter lives in an SU(2) multiplet (for example, higgsinos and winos in supersymmetry) everything is fixed by gauge interactions. For wino-like dark matter, the preferred final state is $W^+W^-$. The lightest neutralino is quasi-degenerate with the lightest chargino, with mass splittings on the order of a fraction of a GeV, and both lie at a scale close to $M_2$, the corresponding soft supersymmetry breaking mass. Coannihilation plays obviously a very significant role, and coannihilation is, here, of the ``symbiotic'' type (charginos pair-annihilate quite efficiently). The resulting pair-annihilation cross section is, approximately \cite{welltempered}
$$
\langle\sigma v\rangle_{\tilde W}\simeq\frac{3g^4}{16\ \pi\ M_2^2}
$$
and the resulting thermal relic abundance is 
$$
\Omega_{\tilde W}h^2\simeq0.1\ \left(\frac{M_2}{2.2\ {\rm TeV}}\right)^2,
$$
implying that thermal winos must weigh about 2.2 TeV. Higgsinos, instead, come in a set of two neutralinos and a chargino with again small mass splittings and at a mass scale around $\mu$. The dominant annihilation final states are $W^+W^-$ and $ZZ$, and the pair annihilation and thermal relic densities are given by
$$
\langle\sigma v\rangle_{\tilde H}\simeq\frac{g^4}{512\ \pi\ \mu^2}\left(21+3\tan^2\theta_W+11\tan^4\theta_W\right)
$$
and
$$
\Omega_{\tilde H}h^2\simeq0.1\ \left(\frac{\mu}{1\ {\rm TeV}}\right)^2
$$
indicating that thermal higgsinos like to weigh about a TeV.

\subsection*{Indirect Detection: Warm-up lap}

Time to charge ahead on indirect detection. The name of the game is to get ``enough'' number counts, in symbols:
$$
N=\phi_\chi\cdot A_{\rm eff}\cdot T_{\rm exp},
$$
where: 
\begin{itemize}
\item $\phi_\chi$ indicates the relevant dark matter-induced event rate, such as the flux of a certain type of Standard Model particle, and has units ${\rm cm}^{-2}\cdot {\rm s}^{-1}$; 
\item $A_{\rm eff}$ is an effective area: good to have in mind some numbers here. For example, in the business of gamma-ray telescopes, the Fermi Large Area Telescope (LAT) has an effective area of about 1 m$^2$, while the top-of-the-line atmospheric Cherenkov (ground based) telescopes, such as H.E.S.S. or MAGIC or VERITAS have effective areas on the order of $10^5\ {\rm m}^2$; the relevant numbers for the two key antimatter satellites are $\sim0.01$ m$^2$ for Pamela and $\sim0.1$ m$^2$ for AMS-02; finally, if you're asked to quote a number for high-energy neutrino telescopes, mention IceCube and mumble 1 km$^2$.
\item $T_{\rm exp}$ indicates the relevant ``exposure time'': for satellites this is on the order of a year, which as you know is exactly $\pi\times 10^7$ s; for typical ground-based telescopes you can perhaps count on about 100h, or about $10^5$ s, while balloon experiments have a typical exposure time of the order of a week, or about $10^6$ s.
\end{itemize}

To detect a signal we need to fulfill two basic conditions: 
\begin{itemize}
\item[(i)] have some signal events, i.e. $\phi_\chi\cdot A_{\rm eff}\cdot T_{\rm exp}\gg1$

\item[(ii)] have enough signal-to-noise, for example requesting $N_{\rm signal}>(\#\sigma)\sqrt{N_{\rm background}}$.
\end{itemize}

I like to classify astrophysical probes of dark matter into three categories:
\begin{enumerate}
\item {\bf Very indirect}: this category includes effects induced by dark matter on astrophysical objects or on cosmological observations;
\item {\bf Indirect}: I include in this category probes that don't ``trace back'' to the annihilation event, as their trajectories are bent as the particles propagate: charged cosmic rays;
\item {\bf Not-so-indirect}: neutrinos and gamma rays -- we will discuss these guys in the next lecture.
\end{enumerate}

{\bf 1. Effects on Astrophysical Objects}: folks have thought about an amazing variety of possibilities, including:
\begin{itemize}
\item Solar Physics (dark matter can affect the Sun's core temperature, the sound speed inside the Sun,...)
\item Neutron Star Capture, possibly leading to the formation of black holes (notably e.g. in the context of asymmetric dark matter, see e.g. \cite{katrinetal})
\item Supernova and Star cooling (see the excellent book by Georg Rafelt \cite{rafelt})
\item Protostars (e.g. WIMP-fueled population-III stars, available also in Swedish \cite{swedish})
\item Planets warming
\end{itemize}
Given the relevance of global warming to the general public (and to funding agencies), let's make an estimate of this latter effect. The ``capture probability'' for WIMPs is roughly
$$
n_{\rm nucleons}\cdot \sigma_{\chi-N}\cdot R_{\rm planet}\lesssim 10^{-4},
$$
where for the right-hand side I've used $n_{\rm nucleons}\sim N_A/{\rm cm}^3$, the current upper limit on the spin-dependent WIMP-nucleon cross section $\sigma_{\chi-N}\lesssim10^{-37}\ {\rm cm}^2$ and the radius of Uranus, $R\sim3\times 10^9$ cm -- the choice of Uranus is motivated by an anomalous heat observed in the planet, of about $10^{14}$ W. Now, the power produced by dark matter assuming that all of the dark matter mass is converted to heat is
$$
W\sim ({\rm capture\ probability})\cdot\pi R_{\rm planet}\cdot \rho_{\rm DM}\cdot v_{\rm DM}\lesssim 10^{12}\ {\rm W}
$$
which tells us that we fall short by a couple orders of magnitude of explaining Uranus' anomalous heat. Too bad.
\begin{center}
\fbox {
    \parbox{0.8\linewidth}{
    {\bf Exercise \#5:} Estimate the heat produced by dark matter annihilation in the Earth and compare with the accuracy of geothermal models (see also the much, much more refined discussion in \cite{bertoneheat}); how large should the dark matter-nucleon scattering cross section to cause global warming concerns?
    }
}
\end{center}

{\bf 1., cnt'd: Effects on Cosmology}: lots of work here, spanning effects on Big Bang Nucleosynthesis, on the cosmic microwave background, on reionization, on structure formation and many more. I don't even have time to give you a laundry list of all this! Go browse the arXiv and have fun!

{\bf 2. Charged Cosmic Rays}: here, the dark matter ``source term'' is unfortunately tangled with effects of propagation and energy losses of charged cosmic rays on their way to our human detectors. 

Do we expect enough cosmic rays from dark matter annihilation or decay to detect a signal over the background? The ballpark energy density of cosmic rays in the Milky Way is $$\epsilon_{\rm CR}\sim1\ \frac{\rm eV}{{\rm cm}^3}.$$
Let's estimate the energy density in cosmic rays dumped by dark matter annihilation in the Galaxy:
$$
\epsilon_{\rm DM}\sim m_\chi\cdot\langle\sigma v\rangle\cdot n_{\rm DM}^2\cdot T_{\rm MW},
$$
with $m\chi\sim100$ GeV, $\rho_{\rm DM}\sim 0.3 \ {\rm GeV}/{\rm cm}^3$, $\langle\sigma v\rangle\sim3\times 10^{-26}\ {\rm cm}^3/{\rm s}$, and the Milky Way age $T_{\rm MW}\sim 10\times 10^9$ yr, I get 
$$
\epsilon_{\rm DM}\sim 10^{-2}\ \frac{\rm eV}{{\rm cm}^3}.
$$
\begin{center}
\fbox {
    \parbox{0.8\linewidth}{
    {\bf Exercise \#6:} Improve on the estimate above using a Navarro-Frenk-White dark matter density profile and integrating over an appropriate cosmic-ray ``diffusion region'', e.g. a cylindrical slab of half-height 1 kpc and radius 20 kpc.
    }
}
\end{center}
\begin{center}
\fbox {
    \parbox{0.8\linewidth}{
    {\bf Exercise \#7:} Same as Ex.\#6, but for a decaying dark matter particle, find $\epsilon_{\rm DM}(\tau)$ where $\tau$ is the dark matter lifetime. Do you expect to get interesting limits on $\tau$ from this calculation? If yes, please mention me and these lecture notes in the acknowledgements of your forthcoming paper.
    }
}
\end{center}

The estimate above indicates that the contribution of annihilating dark matter to cosmic rays is, at best, subdominant to the observed cosmic ray energy density, but that it could be an ${\cal O}(1\ \%)$ effect. In fact, models of Galactic cosmic rays decently match observation, so this is in some sense good news for dark matter model building! As a result, it is key in this business to target under-abundant species, namely either heavy nuclei or antimatter (for example positrons ($e^+$), antiprotons ($\bar p$), antideuterons $\bar D$,...). Unfortunately, it is quite hard to produce heavy nuclei from dark matter annihilation (that results, in its hadronic part, in a couple of high-energy jets only). Antimatter, on the other hand, is promising; typical dark matter models (exceptions are certain flavors of asymmetric dark matter) are democratic in producing as much matter as antimatter in the annihilation or decay final products.

\begin{figure}[t]
\centering
\begin{minipage}{0.45\textwidth}
\mbox{\hspace*{-5cm}\includegraphics*[scale=0.75]{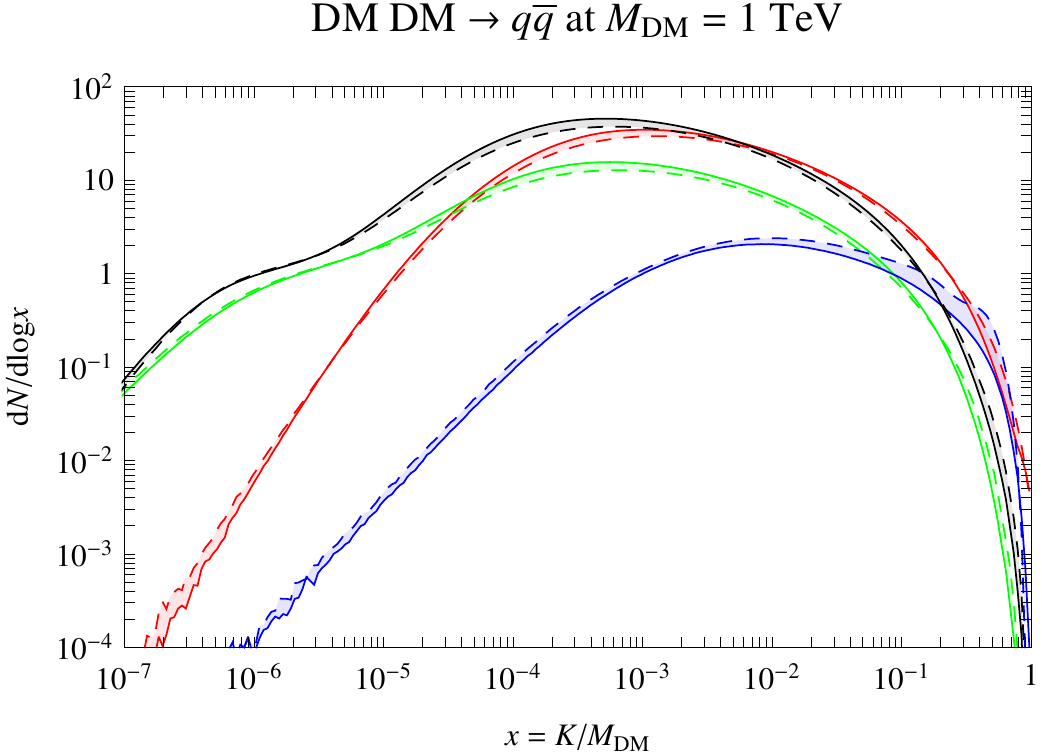}\qquad\includegraphics*[scale=0.75]{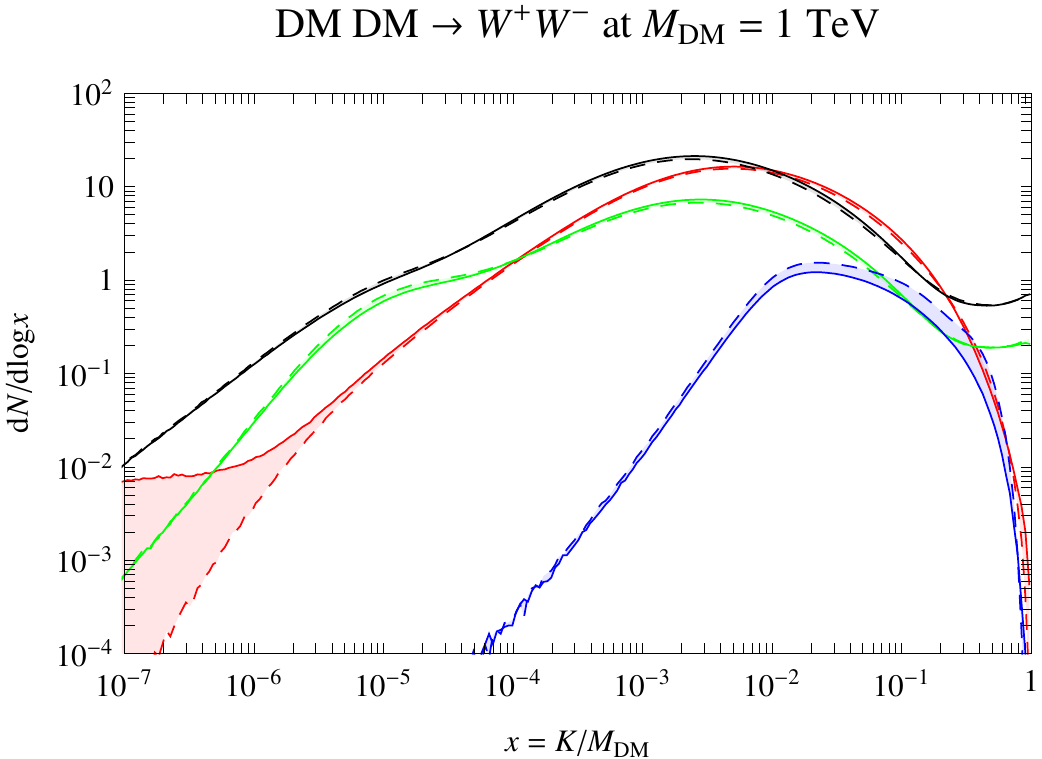}}
\end{minipage}
 \hfill  
\caption{The differential photon (red lines), neutrino (black lines), $e^\pm$ (green lines), $\bar p$ (blue lines) yield from dark matter pair-annihilation into a $q\bar q$ pair (left) and $W^+W^-$ (right). From Ref.~\cite{cirellistrumia}.
\label{fig:cirelli}}
\end{figure}

Figure \ref{fig:cirelli} illustrates the final yield of several particle species resulting from \mbox{$\chi\chi\to q\bar q$} (left) and from \mbox{$\chi\chi\to W^+W^- $} (I took these two nice figures from Ref.~\cite{cirellistrumia}). The red lines indicate photons, the black lines neutrinos, while the green and blue lines indicate $e^\pm$ and $\bar p$, respectively. All of these particle species primarily originate from the hadronization and cascade decays of jets initiated by the final state $q$ and $\bar q$, or directly from the prompt decay modes of the $W$ (notice the green and black lines getting ``horizontal'' at $x=1$, where $x$ is the particles' kinetic energy normalized by the dark matter mass).

\begin{figure}[t]
\centering
\begin{minipage}{0.45\textwidth}
\mbox{\hspace*{-5cm}\includegraphics*[scale=0.45]{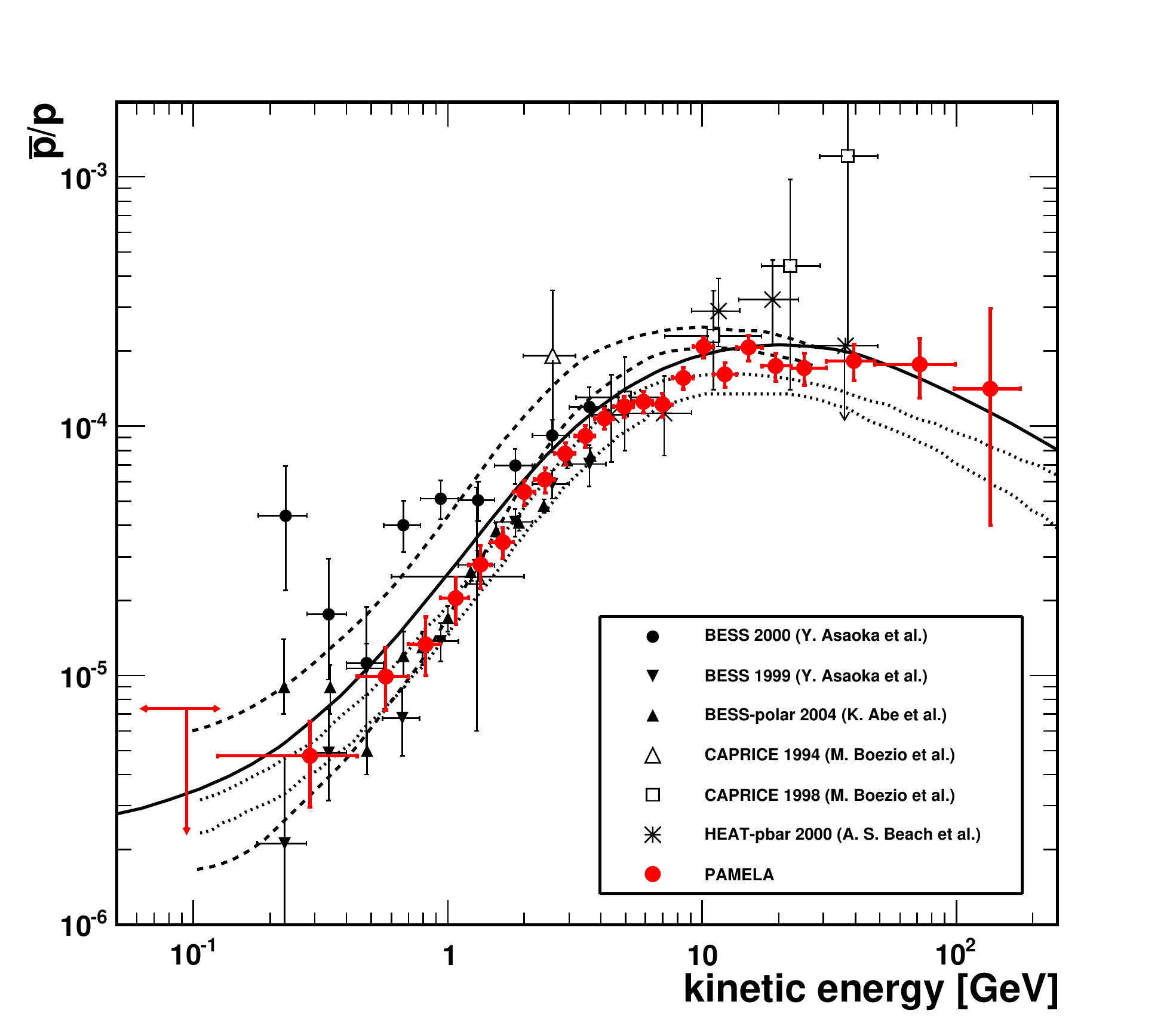}\qquad\includegraphics*[scale=0.41, angle=90]{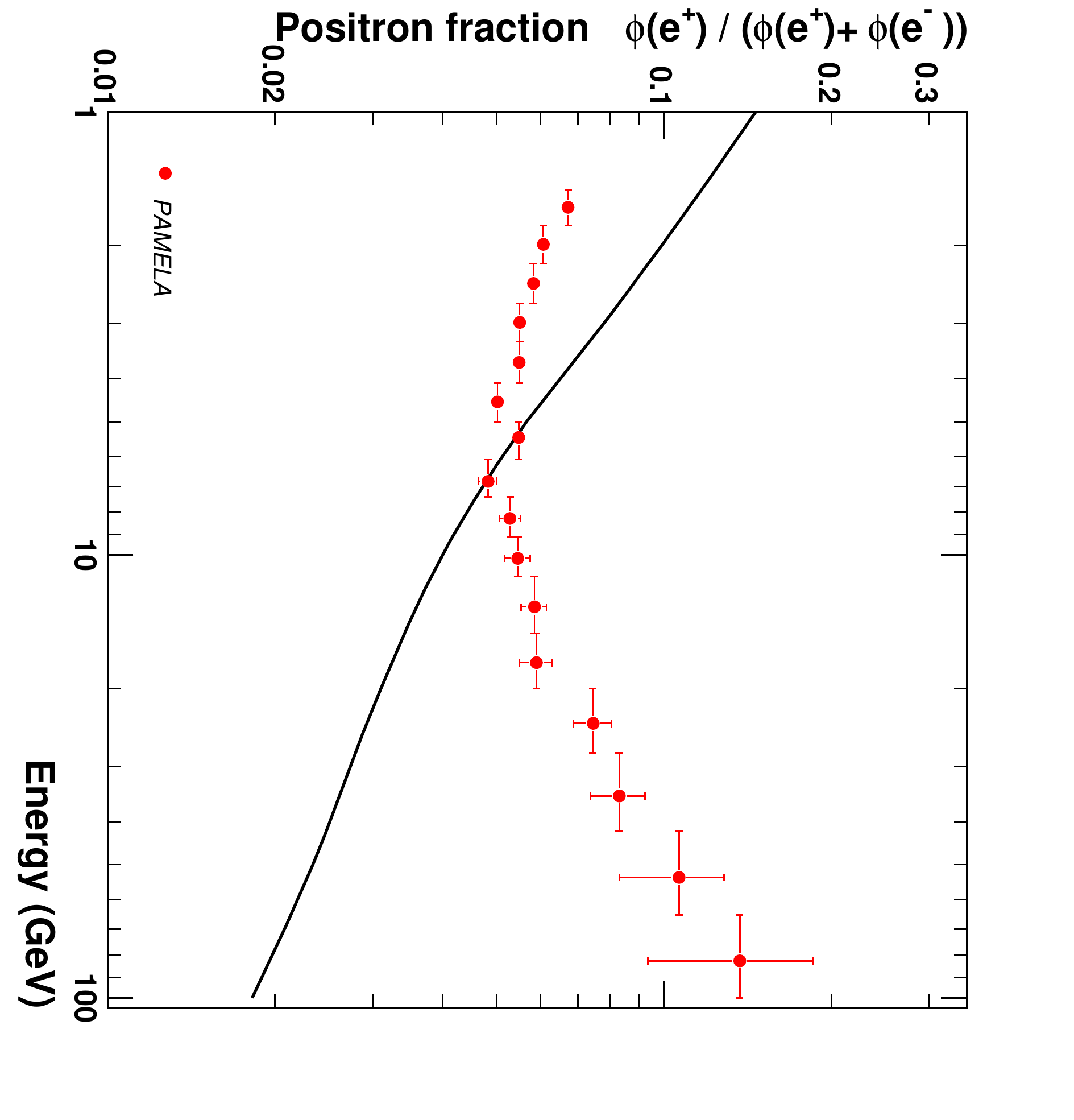}}
\end{minipage}
 \hfill  
\caption{The cosmic-ray antiproton to proton ratio (left, from Ref.~\cite{pamelapbar}) and the positron fraction (right, from Ref.~\cite{pamelapositron}) as measured by the Pamela experiment.
\label{fig:antimatter}\label{fig:pbar}}
\end{figure}

In cosmic rays, antimatter is primarily produced by spallation processes, such as $$p+p\to p+p+\bar p+p$$ where one of the protons in the initial state is a high-energy particle, and the second one is typically an $H^+$ nucleus in the interstellar medium gas, and baryon number conservation forces you to produce at least four nucleons in the final state. The process has a relatively large threshold (if you need a special relativity refresher carry out the two-lines calculation), $E_p\gtrsim 7$ GeV. Now, the spectrum of cosmic rays observed in the Galaxy falls steeply with energy,
$$
\frac{{\rm d}N_{\rm cosmic-ray\ protons}}{{\rm d} E}\ \sim \ E^{-2.7},
$$
so compared to the maximal flux of cosmic-ray protons, observed at $E\sim 0.1$ GeV, antiprotons will be under abundant, at 0.1 GeV, by about a factor 
$$
\frac{\bar p}{p}\sim\left(\frac{0.1}{7.5}\right)^{2.7}\sim10^{-5}.
$$
This is in fact in remarkable agreement with what is observed, see Fig.~\ref{fig:pbar}, left, from Ref.~\cite{pamelapbar}.

There are therefore two effects that make antiprotons an interesting probe of dark matter (that, as fig.~\ref{fig:cirelli} shows, tends to produce low-energy antinucleons): on the one hand there are few ``beam'' particles to produce cosmic-ray antiprotons, since the cosmic-ray proton spectrum falls steeply, and on the other hand the typical kinetic energy inherited by the final state antiproton will be on the same order as the threshold for the process. Indeed, Fig.~\ref{fig:pbar}, left, shows that the $\bar p/p$ ratio peaks right around 10 GeV, a much higher energy than the typical anti nucleon produced by dark matter. These two effects are even more drastic for anti-deuterons (i.e., bound states of $\bar p$ and $\bar n$), for which the key astrophysical background comes from the reaction
$$
\bar D:\quad p+p\to p+p+\bar p+p+\bar n+n$$
that has a threshold of about 17 GeV. In addition, $\bar D$ have a hard time loosing energy by elastic scattering (tertiary population) since the deuteron binding energy is very low, and when hit $\bar D$ tend to disintegrate rather than lose energy! There is a smart idea out there (the proposed satellite is called GAPS \cite{gaps}) to target specifically low-energy antideteurons and to detect them via the peculiar de-excitation X-rays that an atom capturing a $\bar D$ would produce.

\subsection*{How to deal with charged cosmic rays}

How do we model cosmic-ray transport? The most successful framework is provided by the so-called diffusion models (adequate for cosmic-ray energies $E_{\rm CR}\lesssim 10^{17}$ eV). Let us indicate the differential (in energy) number density of cosmic rays with
$$
\frac{{\rm d}n}{{\rm d} E}=\psi\left(\vec x, E,t\right).
$$
The master equation of cosmic-ray diffusion models looks something like this:
\begin{equation}\label{eq:cr}
\frac{\partial}{\partial t}\psi\ =\ D(E)\Delta\psi\ +\ \frac{\partial}{\partial E}\left(b(E)\ \psi\right)\ +\ Q\left(\vec x, E,t\right),
\end{equation}
The first term on the right-hand side describes diffusion, the second one energy losses, and the third includes all possible sources. As always in life and in science, it is possible (and easy!) to add complications -- an incomplete list of popular ones and of the associated recipes is:
\begin{itemize}
\item {\em Cosmic-ray convection}; recipe: add: $\frac{\partial}{\partial z}(v_c\cdot \psi)$.
\item {\em Diffusive re-acceleration}; recipe: add: $\frac{\partial}{\partial p}\ p^2\ D_{pp}\frac{\partial}{\partial p}\frac{1}{p^2}\psi$.
\item {\em Fragmentation and decays}; recipe: add: $-\frac{1}{\tau_{\rm f,d}}\psi$.
\end{itemize}
When dealing with partial differential equations, we all learned in kindergarten that it is crucial to define boundary conditions. A popular choice is free-escape at the boundaries of a ``diffusive region'', whose geometry, for obvious reasons, is typically chosen to be a cylindrical slab, with $$R\sim {\cal O}(1)\times 10\ {\rm kpc},$$ $$h\sim{\cal O}(1)\times 1\ {\rm kpc}.$$ These numbers (very) approximately reflect the distribution of gas and stars in our own Milky Way.

The diffusion coefficient (that in certain models can depend also on position - it more than likely does in reality!) has a dependence on energy (a remnant of the fact that the Larmor radius scales with the particle's momentum!) that can be schematically cast as
$$
D(E)\ \sim\ D_0\left(\frac{E}{E_0}\right)^\delta,\qquad E_0\sim {\rm GeV},\ \ D_0\sim{\rm few}\ \times\ 10^{28}\ \frac{{\rm cm}^2}{\rm s},\quad \delta\sim 0.7.
$$
The parameters entering cosmic ray diffusion are tuned self-consistently to reproduce key observational data, such as stable pure secondary to primary ratios as a function of energy (classic example: boron to carbon, B/C) or unstable secondary to primary ratios, such as ${}^{10}$Be/${^9}$Be. For example, this latter ratio constrains quite severely the height of the diffusion region.

What are the relevant time-scales for the diffusion equations? Two key quantities are the diffusion and the energy loss time scales:
$$
\tau_{\rm diff}\sim\frac{R^2}{D_0}\cdot E^{-\delta},\qquad \tau_{\rm loss}\sim \frac{E}{b(E)},
$$ 
where $R$ is the linear size of the diffusion region, or the relevant time/distance scale for which we want to calculate the typical associated diffusion length (for example, to infer which diffusion length corresponds to the energy loss time scale, we would plug in $R\sim c/\tau_{\rm loss}$). The steady-state diffusion equation (\ref{eq:cr}) can then be re-written as
$$
0\ =\ -\frac{\psi}{\tau_{\rm diff}}\ -\frac{\psi}{\tau_{\rm loss}}+Q,
$$
implying that 
\begin{equation}\label{eq:crapprox}
\psi\sim Q\cdot {\rm min}[\tau_{\rm diff},\ \tau_{\rm loss}].
\end{equation}
Let's see if this makes sense and consider cosmic ray protons and primary and secondary electrons and positrons:
\begin{itemize}
\item If the primary sources of cosmic-ray protons are supernova remnants, and if the injected particles are accelerated via a Fermi mechanism, we expect $$Q\ \sim\ E^{-2}.$$ Energy losses for protons in the GeV-TeV range are relatively inefficient, and typically $\tau_{\rm diff}\ll\tau_{\rm loss}$, therefore Eq.~(\ref{eq:crapprox}) would predict
$$
\psi \ \sim \ E^{-2}\cdot E^{-\delta}\ \sim\ E^{-2.7}
$$
which is in great agreement with observation!
\item For primary electrons, let us suppose that again $Q\sim E^{-2}$ -- for example because the acceleration site is the same as for cosmic-ray protons (not such an unreasonable assumption...). At high energy ($E_e\gg$ GeV) the dominant energy loss mechanisms are inverse-Compton scattering (i.e. the process of a high-energy electron up-scattering an ambient photon -- the inverse of the classic Compton scattering where a high-energy photon up-scatters an electron at rest!) and synchrotron. Both have the same dependence on energy, $\propto E^2$, and the resulting energy loss term reads
$$
b_e(E)\simeq\ b_{\rm IC}^0\left(\frac{u_{\rm ph}}{1\ {\rm eV}/{\rm cm}^3}\right)\cdot E^2\ +\ b_{\rm sync}^0\left(\frac{B}{1\ \mu{\rm G}}\right)\cdot E^2,
$$
where, in units of  $10^{-16}\ {\rm GeV}/{\rm s}$, the constants $$b_{\rm IC}^0\simeq0.76,\qquad b_{\rm sync}^0\simeq 0.025$$ and $u_{\rm ph}$ corresponds to the background radiation energy density and $B$ to the ambient magnetic field. Depending on the size and geometry of the diffusion region, Eq.~(\ref{eq:crapprox}) predicts a break between a low-energy regime where 
$$
\psi_{\rm primary,\ low-energy}\sim\ Q\cdot \tau_{\rm diff}\sim E^{-2}\cdot E^{-\delta}\sim E^{-2.7}
$$
and a high-energy regime where
$$
\psi_{\rm primary,\ high-energy}\sim\ Q\cdot \tau_{\rm loss}\sim E^{-1}\cdot \frac{E}{E^2}\sim E^{-3}.
$$
The general prediction is thus of a {\em broken power-law} with a break corresponding to $\tau_{\rm loss}\sim\tau_{\rm diff}$. This indeed matches observation again! (both directly and indirectly, i.e. from measurements of the secondary radiation produced by cosmic ray $e^\pm$).
\item For secondary electrons and positrons, produced e.g. by the decay of charged pions produced by cosmic-ray proton collisions with protons in the interstellar medium, the source term corresponds to the $Q_p\sim E^{-2.7}$ spectrum found above. The $e^\pm$ spectrum after diffusion and energy losses will then follow the same fate as that of primary particles discussed above: a broken power-law, with a hard low-energy spectrum $\psi_{secondary,\ low-energy}\sim E^{-3.4}$ and a softer high-energy tail due to energy losses, $\psi_{secondary,\ high-energy}\sim E^{-3.7}$. The key point is that, independently of the value of $\delta$ (that, remember, tunes the energy dependence of the diffusion coefficient) and of the primary injection spectrum and of energy, the ratio of secondary to primary species is
$$
\frac{\psi_{e^+}}{\psi_{e^-}}\ \sim\ E^{-\delta}.
$$
The prediction of a declining secondary-to-primary ratio was recently found to be at odds with the observed local positron fraction (see the right panel of figure \ref{fig:pbar}, from Ref.~\cite{pamelapositron}), a fact that spurred much speculation about the nature of the additional positrons responsible for the upturn in the secondary-to-primary ratio.
 
\end{itemize}

There are a couple of special limits in which one can get a simple solution to Eq.~(\ref{eq:cr}) that are worth remembering because they apply to certain interesting physical situations:

\begin{enumerate}
\item {\bf No diffusion}: this case corresponds to physical situations where the energy loss time-scale is much shorter than the diffusion time-scale: the cosmic rays effectively loose their energy before diffusing. In this case, the asymptotic, steady-state (i.e. after time transients) solution to the diffusion equation (\ref{eq:cr}) is
$$
\psi(\vec x,E)\propto\frac{1}{b(E)}\int\ {\rm d}E^\prime\ Q(\vec x,E^\prime).
$$
There are numerous circumstances where this is a relevant approximation: for example, when the system is very large, with its physical size much larger than the typical diffusion length associated with the energy-loss time (for example clusters of galaxies), or when the system is such that the energy loss term is very large (for example, a medium with a very dense radiation fields off of which $e^\pm$ can efficiently inverse-Compton scatter, or with large magnetic field inducing large synchrotron radiation energy losses).
\item {\bf Burst-like injection from a point source at time $t$ in the past}: in this case, the relevant (spherically symmetric) solution, neglecting energy losses, is
\begin{equation}
\label{eq:expsup}
\psi\propto Q\cdot\exp\left(-\left(\frac{r}{r_{\rm diff}}\right)^2\right),
\end{equation}
where $r$ is the source distance, $$r_{\rm diff}\simeq\sqrt{D(E)\cdot t}.$$ 
\end{enumerate}
This second case, a burst-like injection, is especially important in connection with Galactic pulsars as sources of high-energy $e^\pm$, a potential explanation to the anomalous rising positron fraction found by Pamela \cite{pamelapositron} and recently confirmed by the Fermi-LAT \cite{fermiepem}, as pointed out by many (including Yours Truly, who once again is not shy to self-promote his own papers, see e.g. \cite{mypsr}).

What are the requirements on the age and distance of a pulsar that contributes to the Pamela positron anomaly? It is easy to give general arguments: first, the pulsar age must be much shorter than the energy loss time scale for energies as large as about $E_e\sim100$, in order to have at least some energetic $e^\pm$ around! This implies the condition
$$
T_{\rm psr}\ll\tau_{\rm loss}=\frac{E}{b(E)};\ {\rm for}\ E=100\ {\rm GeV}, \tau_{\rm loss}\sim\frac{100}{10^{-16}\cdot 100^2}\ {\rm s}\sim 10^{14}\ {\rm s}\sim 3\ {\rm Myr}.$$
Now, to avoid the exponential suppression of Eq.~(\ref{eq:expsup}) we must have
$$
\sqrt{D(E)\cdot T_{\rm psr}}\gg\ {\rm distance} \to {\rm distance} \ll (3\times10^{28}\cdot 100^{0.7}\cdot 10^{14})^{1/2}\ {\rm cm}\sim 10^{22}\ {\rm cm}\sim 3\ {\rm kpc}.
$$
So our candidate pulsar is younger than about a mega-year and closer than a few kilo-parsec. One would also like the pulsar to have enough power injected in electron-positron pairs, but this condition, for such a nearby object, is usually fulfilled. Interestingly, several pulsar candidates exist within the desired age and distance, including possibly a handful of the newly discovered radio-quiet gamma-ray pulsars detected by Fermi-LAT (see e.g. Ref.~\cite{psrgr}).

As some of you might be aware of, the dark matter annihilation explanation to the Pamela positron fraction anomaly gathered quite a bit of attention (in fact on the order of $10^3$ publications entertain this possibility!). Dark matter as a source of the observed excess high-energy positrons faces various issues, including the following:
\begin{itemize}
\item there is no evidence for an associated antiproton excess, thus the dark matter must preferentially pair-annihilate into non-hadronic final states (it must be ``leptophilic'');
\item diffuse secondary radiation from internal bremsstrahlung and inverse-Compton is not observed;
\item the needed pair-annihilation rate, $$\langle\sigma v\rangle\sim10^{-24}\frac{{\rm cm}^3}{\rm s}\cdot\left(\frac{m_\chi}{100\ {\rm GeV}}\right)^{1.5}$$ is very large for thermal production, and generically leads to unseen gamma-ray or radio emission;
\item a XIII century monk pointed out that ``{\em entia non sunt multiplicanda pr\ae ter necessitatem}'', and the pulsar explanation works just fine to explain the excess positrons.
\end{itemize}
Despite these difficulties, theorists from all over the world (including myself) have proposed models that circumvent all difficulties and show proof that Pamela might have perhaps detected the first non-gravitational signs of dark  matter, providing more and more empirical evidence in favor of the so-called "Redman theorem" \cite{longair}:

\begin{center}
\fbox {
    \parbox{0.85\linewidth}{
    \em Any competent theoretician can fit any given theory to any given set of facts
    }
}
\end{center}

What's next? Well, myself and many others are quite eagerly awaiting results from the AMS-02 payload, successfully deployed and operational on the International Space Station since May 2011. An independent measurement of the positron fraction over an extended energy range (especially in the critical high-energy end, where a cut-off in the positron fraction might indicate a new physics origin!) and with much larger statistics, measurements of various cosmic ray species which will be key to a better understanding of cosmic ray propagation in the Galaxy, and possibly additional information on e.g. anisotropies of the positron arrival direction don't quite keep me awake  as much as my newborn son (actually, not even nearly), but still\ldots

\clearpage

\section*{Lecture 4: not-so-Indirect Detection: Neutrinos and Gamma Rays \label{sec:lecture4}}
\addcontentsline{toc}{section}{Lecture 4: not-so-Indirect Detection: Neutrinos and Gamma Rays}

\subsection*{The tiny neutral ones}

Detecting neutrinos (from an Italian made-up word that indicates the ``tiny neutral one'', with an English made-up plural form) is hard. In fact, despite building km$^3$ size detectors, only two astrophysical neutrino sources have been observed so far: the Sun and Supernova 1987A! The flip side of the coin is that astrophysical backgrounds are evidently quite low (albeit of course cosmic rays produce copious ``atmospheric'' neutrinos as they hit the atmosphere...) if one is to use neutrinos to search for dark matter. The key idea is that dark matter particles can accrue in celestial bodies until large enough densities start fueling a steady rate of annihilation yielding high-energy neutrinos. Neutrinos are pretty much the only thing produced by dark matter annihilation that can escape the core of a celestial body without losing much energy at all, and get all the way out to our km$^3$ size detectors. The best bets are the Sun and the Earth, with the former, turns out, much better  than (although somewhat complementary to) the latter. Let us now make a few estimates for this process, for the case of dark matter capture and annihilation in the Sun.

The dark matter capture rate in the Sun is, roughly
$$
C^\odot\sim\ \phi_\chi\cdot \left(\frac{M_\odot}{m_p}\right)\cdot \sigma_{\chi-p},
$$
with the dark matter flux
$$\phi_\chi\sim n_\chi\cdot v_{\rm DM}=\frac{\rho_{\rm DM}}{m_\chi}\cdot v_{\rm DM}$$ 
the ratio $M_\odot/m_p$ estimating the number of target nucleons in the Sun, and the dark matter-nucleon interaction cross section $\sigma_{\chi-p}$ being bound by current experimental limits:
\begin{eqnarray*}
\sigma_{\chi-p}^{\rm spin\ dependent}&\lesssim&\ 10^{-39}\ {\rm cm}^2,\\
\sigma_{\chi-p}^{\rm spin\ independent}&\lesssim&\ 10^{-44}\ {\rm cm}^2.
\end{eqnarray*}
Plugging in the relevant numbers, I find
$$
C^\odot\sim\frac{10^{23}}{\rm s}\ \left(\frac{\rho_{\rm DM}}{0.3\ {\rm GeV}/{\rm cm}^3}\right)\cdot \left(\frac{v_{\rm DM}}{300\ {\rm km/s}}\right)\cdot \left(\frac{100\ {\rm GeV}}{m_\chi}\right)\cdot \left(\frac{\sigma_{\chi-p}}{10^{-39}\ {\rm cm}^2}\right)
$$
We are interested in the number of dark matter particles in Sun: let's call this number $N$ and write down a differential equation that describes the time evolution of $N(t)$:
$$
\frac{{\rm d}N}{{\rm d}t}=C^\odot-A^\odot[N(t)]^2-E^\odot N(t),
$$
There are various elements I introduced in the otherwise self-explanatory equation above:
\begin{itemize}
\item $E^\odot$ describes the ``evaporation'' of dark matter particles, something that happens if the particles have a (thermal) velocity comparable with the celestial body's escape velocity. Let's quickly estimate this effect. For the Sun $$v^\odot_{\rm esc}\simeq1156\ \frac{\rm km}{\rm s}\sim3\times 10^{-3}\ c$$ while the Sun's core temperature (the dark matter particles sink to the center after multiple scattering inside the Sun) is $$T^\odot_{\rm core}\sim10^7\ K\sim1\ {\rm keV}\sim m_\chi\cdot v_\chi^2.$$ This gives, for the typical dark matter thermal velocities in the core of the Sun
$$
v_\chi\sim c\cdot \left(\frac{1\ {\rm keV}}{m_\chi}\right)^{1/2}\gtrsim v_{\rm esc}^\odot \rightarrow m_\chi\lesssim 0.1\ {\rm GeV}.
$$
Bottom line: for dark matter particles in the ``preferred'' WIMP mass range we can safely neglect evaporation.

\begin{center}
\fbox {
    \parbox{0.8\linewidth}{
    {\bf Exercise \#8:} Estimate evaporation in the case of the Earth: what is the relevant dark matter particle mass range for which evaporation matters in the Earth?
    }
}
\end{center}

\item The annihilation rate $$A^\odot\simeq\frac{\langle\sigma v\rangle}{V_{\rm eff}},$$ where $V_{\rm eff}$ is an effective volume which depends on where WIMPs live inside the Sun; let us use the following (reasonable) guess for the density profile of the sunk WIMPs in the Sun:
$$
n(r)=n_0\ \exp\left(-\frac{m_\chi\phi_{\rm grav}(r)}{T^\odot}\right).
$$
We can choose to estimate $V_{\rm eff}$ by identifying an effective radius $R_{\rm eff}$ corresponding to the condition
$$
\frac{m_\chi\ \phi_{\rm grav}(R_{\rm eff})}{T^\odot}\simeq1\quad \rightarrow\quad T^\odot\simeq\frac{G_N\rho^\odot\frac{4\pi}{3}R_{\rm eff}^3m_\chi}{R_{\rm eff}}
$$
$$
\rightarrow R_{\rm eff}\sim 10^9\ {\rm cm}\ \left(\frac{m_\chi}{100\ {\rm GeV}}\right)^{1/2}\quad\rightarrow\quad V_{\rm eff}\sim 10^{28}\ {\rm cm}^3\ \left(\frac{m_\chi}{100\ {\rm GeV}}\right)^{3/2}
$$
Remember that the Sun's radius is approximately $R^\odot\sim7\times 10^{10}$ cm, so this radius is smaller than the Sun's radius for reasonably light WIMPs. 
\end{itemize}
Neglecting evaporation, even I can solve the differential equation above, and calculate the quantity we are really interested in, the annihilation rate in the Sun:
$$
\Gamma_A=\frac{1}{2}A^\odot[N(t^\odot)]^2=\frac{C^\odot}{2}\Big[\tanh (\sqrt{C^\odot A^\odot}\ t^\odot)\Big]^2,
$$
with $$t^\odot\sim4.5\ {\rm Byr}\sim 10^{17}\ {\rm s}$$ the Sun's age (not to be confused with the Sun's core temperature $T^\odot$!). One thing we learn from the solution above is that equilibrium between capture and annihilation is reached if
$$
t^{\rm eq}\equiv\frac{1}{\sqrt{C^\odot A^\odot}}\ll t^\odot.
$$
Do we expect equilibrium or not, for nominal WIMP parameters? Yes, we do! Let's plug in the numbers and convince ourselves of this fact: first, let's find the required annihilation rate for equilibrium
$$
C^\odot\sim 10^{23}\ {\rm s}^{-1}\ \left(\frac{\sigma_{\chi-p}}{10^{-39}\ {\rm cm}^2}\right),
$$
$$
A^\odot_{\rm eq}\gg\frac{1}{(t^\odot)^2\ C^\odot}=\frac{1}{10^{34}\cdot 10^{23}\ {\rm s}}\sim 10^{-57}\ {\rm s}^{-1}.
$$
Now, for vanilla WIMP dark matter
$$
A^\odot=3\times 10^{-54}\ {\rm s}^{-1} \ \left(\frac{\langle\sigma v\rangle}{3\times 10^{-26}\ {\rm cm}^3/{\rm s}}\right)
$$
so equilibrium is reached for $\sigma_{\chi-p}$ as small as about $10^{-41}\ {\rm cm}^2$. 
\begin{center}
\fbox {
    \parbox{0.8\linewidth}{
    {\bf Exercise \#9:} Re-do this calculation for the case of the Earth and find the critical dark matter-nucleon scattering cross section for equilibrium; note that the relevant scattering cross section in the Earth is spin-independent (as the Earth is mostly made of spin-0 Iron nuclei): do you then expect the equilibrium condition to hold for the flux of neutrinos from the center of the Earth?
    }
}
\end{center}
If equilibrium is achieved, then 
$$
\Gamma_A\simeq \frac{C^\odot}{2}
$$
and we don't care about the pair annihilation cross section (a unique case in the business of indirect dark matter detection!), while we only care about the cross section for dark matter capture. The resulting flux of neutrinos of flavor $f$ will then be
$$
\frac{{\rm d}N_{\nu_f}}{{\rm d}E_{\nu_f}}=\frac{C^\odot}{8\pi(D^\odot)^2}\ \left(\frac{{\rm d}N_{\nu_f}}{{\rm d}E_{\nu_f}}\right)_{\rm inj}
$$
where the last factor with the subscript ``inj'' is the ``injection'' spectrum of neutrinos per annihilation. Effects that complicate this discussion include neutrino oscillation, absorption of neutrinos in the Sun, and many others that a few smart people out there have already kindly worked out for you.

The final step is to count the number of events we expect at IceCube or at any other mega-neutrino-detector (these detectors are fundamentally arrays of photomultipliers reading Cherenkov light from muons produced by $\nu_\mu$ charged-current interactions):
$$
N_{\rm events}=\int {\rm d}E_{\nu_\mu}\int {\rm d}y\ \left(A_{\rm eff}\cdot\frac{{\rm d}N_{\nu_\mu}}{{\rm d}E_{\nu_\mu}}\cdot\frac{{\rm d}\sigma}{\rm d y}(E_{\nu_\mu,\ y})\cdot\left(R_\mu(E_{\nu_\mu})\right)\right),
$$
where $y$ indicates the $\nu_\mu$ energy fraction transferred to the $\mu$ in the charged current interaction, ${{\rm d}\sigma}/{\rm d y}$ is the relevant cross section for charged current interactions, and the last factor $R_\mu$ indicates the muon range in the relevant material the detector lives in (for example, Antarctic ice for IceCube).

The most promising dark matter pair annihilation final states in this business are those producing a ``hard'' spectrum of muon neutrinos, i.e. energetic neutrinos. These by all means include $W^+W^-$ and $ZZ$ pairs, that dump out prompt muon neutrinos from the leptonic decay modes of the gauge bosons; luckily, for example in supersymmetry, these are exactly the preferred final states for wino- and higgsino-like lightest neutralinos. 

The typical flux sensitivity threshold we want to hit to get an interesting signal is about hundreds of muons per km-squared per year, and the typical energy thresholds are 100 GeV for IceCube, which is improved down to 10 GeV for DeepCore and that could go down to the order of a GeV for the further thickly instrumented portion of the detector to be named PINGU.

\subsection*{Light from Dark Matter}

There are two key ways to get light out of dark matter:
\begin{itemize}
\item[(i)] {\em Prompt} photons from the annihilation or decay event, and
\item[(ii)] {\em Secondary} photons from radiative processes associated with stable, charged particles produced by the dark matter annihilation or decay event (in practice, the most important ones are electrons and positrons)
\end{itemize}

Prompt photons are produced either by the two-photon decay of neutral pions $\pi^0\to\gamma\gamma$ dumped by the hadronization chain of strongly interacting annihilation products, or by internal bremsstrahlung off of charged particles in the intermediate or final state; this second contribution is typically ``harder'', i.e. more energetic, than the first one. Gamma rays from neutral pion decay have the nice spectral feature that I ask you to derive in the next exercise\footnote{I remember this problem well, as it was asked to me during my PhD entrance exam by my advisor-to-be!}.

\begin{center}
\fbox {
    \parbox{0.8\linewidth}{
    {\bf Exercise \#10:} Show that, independent of the $\pi^0$ spectrum, the differential spectrum of gamma rays resulting from $\pi^0\to\gamma\gamma$, ${\rm d}N_\gamma^{\pi^0}/{\rm d}E_\gamma$ is symmetric around $E_\gamma=m_\pi/2$ on a log scale in energy.
        }
}
\end{center}

Secondary photons originate as the counterpart of the key energy loss processes for electrons and positrons we discussed in the previous lecture: inverse-Compton and synchrotron. To qualitatively understand the features of inverse-Compton emission, it is useful to commit to memory the formula for the average energy $\langle E_0^\prime\rangle$ of the up-scattered photon (with an original initial energy $E_0$) as a function of the Lorentz factor $\gamma_e=E_e/m_e$ of the impinging high-energy electron:
$$
\langle E_0^\prime\rangle\sim\frac{4}{3}\gamma_e^2\ E_0.
$$
The relevant numbers for $E_0$ are as follows:
\begin{eqnarray*}
&&{\rm CMB}: E_0\sim2\times 10^{-4}\ {\rm eV}\\
&&{\rm starlight}: E_0\sim1\ {\rm eV}\\
&&{\rm dust}: E_0\sim0.01\ {\rm eV}
\end{eqnarray*}
so for a typical electron-positron injection energy from dark matter $$E_e\sim\frac{m_\chi}{10}\rightarrow\gamma_e\sim2\times 10^4\left(\frac{m_\chi}{100\ {\rm GeV}}\right)$$
and $$E^\prime_{\rm CMB}\sim10^5\ {\rm eV}\left(\frac{m_\chi}{100\ {\rm GeV}}\right)^2.$$
Inverse-Compton emission from dark matter therefore produces hard X-ray photons in the hundreds of keV range. This is great news, as a brand new NASA telescope, NuSTAR, is looking at the sky exactly in that energy range \cite{nustar}! The inverse-Compton light from starlight and dust falls, instead, in the low-energy gamma-ray regime.

\begin{figure}[t]
\centering
\begin{minipage}{0.45\textwidth}
\mbox{\hspace*{-5cm}\includegraphics*[scale=0.315]{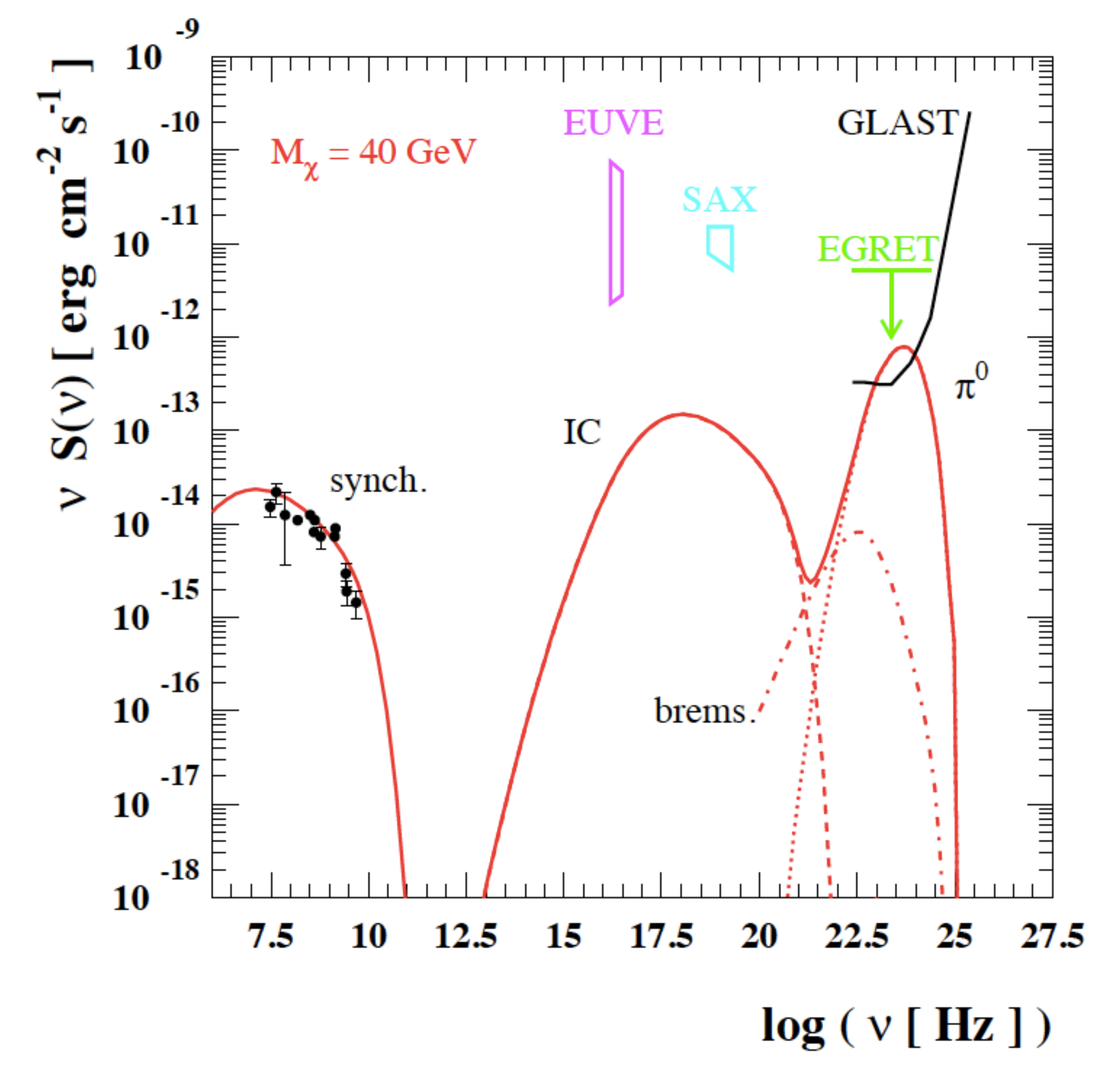}\ \includegraphics*[scale=0.38]{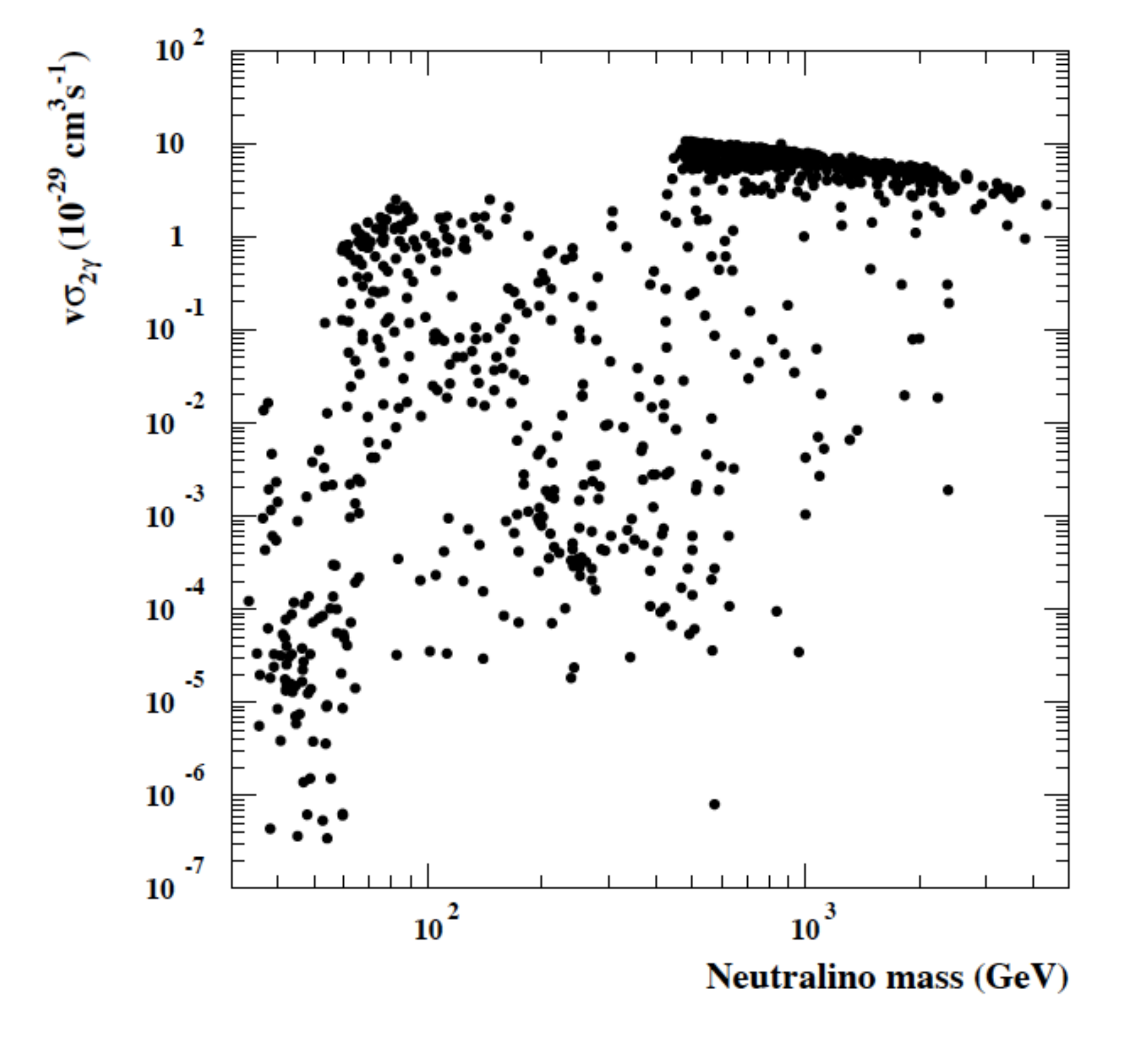}}
\end{minipage}
 \hfill  
\caption{Left: The multi-wavelength emission spectrum from the pair-annihilation of a dark matter particle with $m_\chi=40$ GeV in the Coma cluster of galaxies, from Ref.~\cite{colacoma}. Right: the pair annihilation cross section into two photons for MSSM neutralinos, from Ref.~\cite{ullio}.  \label{fig:coma}}
\end{figure}

In the monochromatic approximation, synchrotron emission peaks at
$$
\frac{\nu_{\rm sync}}{\rm MHz}\simeq2\cdot\left(\frac{E_e}{\rm GeV}\right)\left(\frac{B}{\mu{\rm G}}\right)^{1/2}
$$
and the synchrotron power scales like $B^2$. Dark matter annihilation thus produces a rich, multi-wavelength emission spectrum that goes well beyond the gamma-ray band. An example of the spectrum expected e.g. from the nearby Coma cluster of galaxies is shown in Fig.~\ref{fig:coma}, left, from Ref.~\cite{colacoma}. Note that the various secondary emission peaks appear exactly where the formulae above would predict them to be!

While secondary emission is always present, it involves the additional steps of accounting for the diffusion and energy losses of the $e^\pm$ produced by dark matter annihilation. Prompt gamma-ray emission, on the other hand, is simpler, and it only involves identifying a dark matter structure and a particle dark matter model; we will thus here make a few estimates for this prompt emission only, which for nominal WIMPs produces photons in the gamma-ray energy range. Also, for definiteness we will talk about dark matter annihilation - dark matter decay is even simpler!

What are the optimal targets and the expected detection rates for gamma ray searches for dark matter? The flux of photons produced by dark matter annihilation from a given direction $\psi$ in the sky and from within a solid angle $\Delta\Omega$ is
$$
\phi_\gamma=\frac{\Delta\Omega}{4\pi}\Big\{\frac{1}{\Delta\Omega}\int{\rm d}\Omega\int{\rm d}l(\psi)\ \left(\rho_{\rm DM}\right)^2\Big\}\frac{\langle\sigma v\rangle}{2m_\chi^2}\frac{{\rm d}N_\gamma}{{\rm d}E_\gamma},
$$
where the last factor is, in fact, a sum of the prompt gamma-ray spectrum from every annihilation final state $f$:
$$
\frac{{\rm d}N_\gamma}{{\rm d}E_\gamma}=\sum_f\ \frac{{\rm d}N^f_\gamma}{{\rm d}E_\gamma},
$$
and where the term in curly brackets is often referred to as the ``J-factor'', and is a function of the solid angle $\Delta\Omega$ and of course of the direction in the sky, $J=J(\Delta\Omega,\psi)$, and carries units of ${\rm GeV}^2/{\rm cm}^5$. The solid angle $\Delta\Omega$ should be optimized for a given gamma-ray detector and for a given target, field of view and angular resolution to maximize (typically) the signal to noise. It turns out that the relevant solid angles correspond to an angular extent of about one degree, or $\Delta\Omega\sim 10^{-3}$ sr for the Fermi-LAT at an energy of about a GeV, down to an angular extent of 0.1 degrees, or $\Delta\Omega\sim10^{-5}$ sr for ACT, or for Fermi in the high-energy regime.

Let me give you a ``laundry list'' of potential interesting targets to search for a gamma-ray signal from dark matter; for most of these targets the ``$J$ factor'' is approximately the same for a solid angle corresponding to 1 deg or 0.1 deg:
\begin{enumerate}
\item Dwarf Spheroidal Galaxies
\begin{itemize}
\item Draco, $J\sim10^{19}\ {\rm GeV}^2/{\rm cm}^5$, $\pm$ a factor 1.5;
\item Ursa Minor, $J\sim10^{19}\ {\rm GeV}^2/{\rm cm}^5$, $\pm$ a factor 1.5;
\item Segue, $J\sim10^{20}\ {\rm GeV}^2/{\rm cm}^5$, $\pm$ a factor 3
\end{itemize}
\item Local Milky-Way-like galaxies
\begin{itemize}
\item M31, $J\sim10^{20}\ {\rm GeV}^2/{\rm cm}^5$
\end{itemize}
\item Local clusters of galaxies
\begin{itemize}
\item Fornax, $J\sim10^{18}\ {\rm GeV}^2/{\rm cm}^5$
\item Coma, $J\sim10^{17}\ {\rm GeV}^2/{\rm cm}^5$
\item Bullet, $J\sim10^{14}\ {\rm GeV}^2/{\rm cm}^5$
\end{itemize}
\item Galactic center
\begin{itemize}
\item $0.1^\circ$: $J\sim10^{22}\ldots\ 10^{25}\ {\rm GeV}^2/{\rm cm}^5$
\item $1^\circ$: $J\sim10^{22}\ldots\ 10^{24}\ {\rm GeV}^2/{\rm cm}^5$
\end{itemize}
\end{enumerate}

To have a detection, we need to have enough photon counts, possibly a lot:
$$
N_\gamma\sim\int_{E_\gamma\ {\rm range}}\ {\rm d}E_\gamma\ \phi_\gamma\cdot A_{\rm eff}(E_\gamma)\cdot T_{\rm obs}
$$
The following table gives a rule of thumb for the relevant energy ranges, effective areas and observing time for current and future gamma-ray observatories:

\begin{table}[h]
\begin{center}
\begin{tabular}{|c|c|c|c|}
\hline
 & Fermi-LAT & H.E.S.S. & CTA\\
 \hline
 $E_{\rm \gamma}$ range & 0.1 to 300 GeV & 0.1 to 10 TeV & 10 GeV to 10 TeV \\
 $A_{\rm eff}$ & $\sim 1\ {\rm m}^2$ & $\sim 10^5\ {\rm m}^2$ & $\sim 10^6\ {\rm m}^2$ \\ 
 $T_{\rm obs}$ & $\sim 10^8$ s & $\sim 10^6$ s &$\sim 10^6$ s \\
\hline
\end{tabular}
\end{center}
\label{tab:paramgr}
\end{table}%

It is instructive to calculate the minimal $J$ factor needed to get at least {\em some} gamma-ray signal from dark matter. Consider for example Fermi-LAT: over the LAT energy range, typically
$$
\int\ {\rm d}E_\gamma\ \frac{{\rm d}N_\gamma}{{\rm d}E_\gamma}\ \sim\ \frac{m_\chi}{\rm GeV},
$$
so that
$$
\phi_\gamma=\left(\Delta\Omega\cdot J\right)\frac{1}{8\pi}\frac{\langle\sigma v\rangle}{m_\chi^2}\cdot m_\chi\sim\ 10^{-32}\frac{1}{{\rm cm}^2\ {\rm s}}\left(\frac{J}{{\rm GeV}^2/{\rm cm}^5}\right)
$$
and
$$
N_\gamma\sim A_{\rm eff}\cdot T_{\rm obs}\cdot \phi_\gamma\sim\ 10^{-20}\ \frac{J}{{\rm GeV}^2/{\rm cm}^5},
$$
where we put in the nominal values for the effective area and observing time as in the Table above, so, we want $$J\gtrsim 10^{20}\ {\rm GeV}^2/{\rm cm}^5.$$
This is a bit bigger than the individual dwarf spheroidal galaxies' $J$ factors quoted above. In fact, combining observations of all (non-detected) dwarf galaxies gives one of the tightest (in my personal opinion {\em the} tightest) constraints to date on the dark matter pair-annihilation rate as a function of mass: dwarf galaxies are a virtually background free target, with $$J^{\rm tot}\sim {\rm few}\times10^{20}\ {\rm GeV}^2/{\rm cm}^5,$$
therefore the resulting limits \cite{dsphlim} are
$$\langle\sigma v\rangle_{\rm lim}\ \sim\  3\times 10^{-26}\ \frac{{\rm cm}^3}{\rm s}\ \left(\frac{30\ {\rm GeV}}{m_\chi}\right).
$$
A clear-cut signal of dark matter annihilation that could be detected with gamma-ray detectors is the close-to-monochromatic line from the direct annihilation $$\chi\chi\ \to\ \gamma\gamma.$$ Since $E_\chi\sim m_\chi$, $E_\gamma\simeq m_\chi$ and the line is almost monochromatic. Dark matter particles, being dark, ought better not directly couple to photons, and the naive expectation is that the $\gamma\gamma$ amplitude be loop-suppressed, i.e.
$$
\frac{\langle\sigma v\rangle_{\gamma\gamma}}{\langle\sigma v\rangle_{\rm tot}}\sim\frac{\alpha^2}{16\pi^2}.
$$
This naive estimate gives the correct ballpark over a wide range of parameters in the MSSM, as fig.~\ref{fig:coma}, right (taken from Ref.~\cite{ullio}), illustrates (but things can get a lot worse, and not much better, than the naive estimate!). As you might have heard, a monochromatic line at 130 (or perhaps 135) GeV might indeed be present in the Fermi LAT data \cite{weniger, sufink}, although a solid confirmation (possibly only with the new generation LAT analysis software Pass 8) that this is not an instrumental effect, and that the signal is indeed as statistically robust as it (frankly very much) looks, has yet to come. Stay tuned. Clearly, very exciting days lie ahead!

\newpage
\section*{Acknowledgments}
I would like to thank Elena Pierpaoli for inviting me to lecture at TASI 2012. It was fun, and it was great to meet so many, promising young fellas. Many thanks to the local organizing committee and to the administrative staff as well. Thanks to Eric Carlson, Jonathan Cornell, Jonathan Kozaczuk and Tim Linden for carefully proof-reading and for their feedback on this manuscript. My research is partly supported by the Department of Energy, under grant DE-FG02-04ER41286. This manuscript employs a Latex skeleton I took from Peter Skands' TASI proceedings, arXiv 1207.2389 (it looked too good to pass!!).

\addcontentsline{toc}{section}{References}
\bibliographystyle{utphys}
\bibliography{tasi-profumo}

\providecommand{\href}[2]{#2}\begingroup\raggedright\begin{thebibliography}{100}

\bibitem{KT}
 E.~W.~Kolb, (Ed.) and M.~S.~Turner, (Ed.),
  {\em ``The Early Universe. Reprints,''}
  REDWOOD CITY, USA: ADDISON-WESLEY (1988) 719 P. (FRONTIERS IN PHYSICS, 70)

\bibitem{dodelson}
S.~Dodelson,
  {\em ``Modern cosmology,''}
  Amsterdam, Netherlands: Academic Pr. (2003) 440 p
  
\bibitem{coswikmcc}
R. Coswik and J. McClelland,  Phys.\ Rev.\ Lett.\  {\bf 29}, 669 (1972). According to Ref.~[1], this limit was first derived by G. Gerstein and Ya. B. Zel'dovich, Zh. Eksp. Teor. Fiz. Pis'ma Red. {\bf 4}, 174 (1966).

\bibitem{wimpless}
J.~L.~Feng and J.~Kumar,
  Phys.\ Rev.\ Lett.\  {\bf 101}, 231301 (2008)
  [arXiv:0803.4196 [hep-ph]].

\bibitem{partialwave}
 K.~Griest and M.~Kamionkowski,
  Phys.\ Rev.\ Lett.\  {\bf 64}, 615 (1990).

\bibitem{leeweinberg}
B.~W.~Lee and S.~Weinberg,
  Phys.\ Rev.\ Lett.\  {\bf 39}, 165 (1977).

\bibitem{baltzrev}
 E.~A.~Baltz,
  eConf C {\bf 040802}, L002 (2004)
  [astro-ph/0412170].

\bibitem{relicsusyref}
S.~Profumo,
  Phys.\ Rev.\ D {\bf 78}, 023507 (2008)
  [arXiv:0806.2150 [hep-ph]].

\bibitem{maxphase}
 C.~Wainwright and S.~Profumo,
  Phys.\ Rev.\ D {\bf 80}, 103517 (2009)
  [arXiv:0909.1317 [hep-ph]].

\bibitem{gondologelmini}
 P.~Gondolo and G.~Gelmini,
  Nucl.\ Phys.\ B {\bf 360}, 145 (1991).

\bibitem{gs}
K.~Griest and D.~Seckel,
  Phys.\ Rev.\ D {\bf 43}, 3191 (1991).
  
\bibitem{gondoloedsjo}
 J.~Edsjo and P.~Gondolo,
  Phys.\ Rev.\ D {\bf 56}, 1879 (1997)
  [hep-ph/9704361].
  
\bibitem{uedreview}
D.~Hooper and S.~Profumo,
  Phys.\ Rept.\  {\bf 453}, 29 (2007)
  [hep-ph/0701197].
  
\bibitem{staudeath}
M.~Citron, J.~Ellis, F.~Luo, J.~Marrouche, K.~A.~Olive and K.~J.~de Vries,
  arXiv:1212.2886 [hep-ph].
  
\bibitem{salati}
 P.~Salati,
  Phys.\ Lett.\ B {\bf 571}, 121 (2003)
  [astro-ph/0207396].

\bibitem{myquint}
 S.~Profumo and P.~Ullio,
  JCAP {\bf 0311}, 006 (2003)
  [hep-ph/0309220].
  
\bibitem{buckprofumo}
 M.~R.~Buckley and S.~Profumo,
  Phys.\ Rev.\ Lett.\  {\bf 108}, 011301 (2012)
  [arXiv:1109.2164 [hep-ph]].

\bibitem{bringmann}
 T.~Bringmann,
  New J.\ Phys.\  {\bf 11}, 105027 (2009)
  [arXiv:0903.0189 [astro-ph.CO]].
  
\bibitem{kamionsmallscale}
 S.~Profumo, K.~Sigurdson and M.~Kamionkowski,
  Phys.\ Rev.\ Lett.\  {\bf 97}, 031301 (2006)
  [astro-ph/0603373].

\bibitem{cornell}
J.~M.~Cornell and S.~Profumo,
  JCAP {\bf 1206}, 011 (2012)
  [arXiv:1203.1100 [hep-ph]].
  
  
\bibitem{welltempered}
 N.~Arkani-Hamed, A.~Delgado and G.~F.~Giudice,
  Nucl.\ Phys.\ B {\bf 741}, 108 (2006)
  [hep-ph/0601041].
  
\bibitem{katrinetal}
 S.~D.~McDermott, H.~-B.~Yu and K.~M.~Zurek,
  Phys.\ Rev.\ D {\bf 85}, 023519 (2012)
  [arXiv:1103.5472 [hep-ph]].
  
\bibitem{rafelt}
G.~G.~Raffelt,
  {\em ``Stars as laboratories for fundamental physics: The astrophysics of neutrinos, axions, and other weakly interacting particles,''}
  Chicago, USA: Univ. Pr. (1996) 664 p
  
\bibitem{swedish}
  D.~Spolyar, K.~Freese, P.~Gondolo, A.~Aguirre, P.~Bodenheimer, J.~A.~Sellwood and N.~Yoshida,
  PoS IDM {\bf 2008}, 077 (2008)
  [arXiv:0901.4574 [astro-ph.CO]].
  
\bibitem{bertoneheat}
 G.~D.~Mack, J.~F.~Beacom and G.~Bertone,
  Phys.\ Rev.\ D {\bf 76}, 043523 (2007)
  [arXiv:0705.4298 [astro-ph]].
  
\bibitem{cirellistrumia}
M.~Cirelli, G.~Corcella, A.~Hektor, G.~Hutsi, M.~Kadastik, P.~Panci, M.~Raidal and F.~Sala {\it et al.},
  JCAP {\bf 1103}, 051 (2011)
  [Erratum-ibid.\  {\bf 1210}, E01 (2012)]
  [arXiv:1012.4515 [hep-ph]].
  
\bibitem{pamelapbar}
O.~Adriani {\it et al.}  [PAMELA Collaboration],
  Phys.\ Rev.\ Lett.\  {\bf 105}, 121101 (2010)
  [arXiv:1007.0821 [astro-ph.HE]].
  
\bibitem{pamelapositron}
O.~Adriani {\it et al.}  [PAMELA Collaboration],
  Nature {\bf 458}, 607 (2009)
  [arXiv:0810.4995 [astro-ph]].
  
\bibitem{gaps}
 T.~Aramaki, S.~E.~Boggs, W.~W.~Craig, H.~Fuke, F.~Gahbauer, C.~J.~Hailey, J.~E.~Koglin and N.~Madden {\it et al.},
  Adv.\ Space Res.\  {\bf 46}, 1349 (2010).
  
\bibitem{fermiepem}
 M.~Ackermann {\it et al.}  [Fermi LAT Collaboration],
  Phys.\ Rev.\ Lett.\  {\bf 108}, 011103 (2012)
  [arXiv:1109.0521 [astro-ph.HE]].
  
\bibitem{mypsr}
S.~Profumo,
  Central Eur.\ J.\ Phys.\  {\bf 10}, 1 (2011)
  [arXiv:0812.4457 [astro-ph]];
D.~Grasso {\it et al.}  [FERMI-LAT Collaboration],
  Astropart.\ Phys.\  {\bf 32}, 140 (2009)
  [arXiv:0905.0636 [astro-ph.HE]].

\bibitem{psrgr}
L.~Gendelev, S.~Profumo and M.~Dormody,
  JCAP {\bf 1002}, 016 (2010)
  [arXiv:1001.4540 [astro-ph.HE]].

\bibitem{longair}
 M.~S.~Longair, (ed.),
  {\em ``High-energy astrophysics. Vol. 1: Particles, photons and their detection,''}
  Cambridge, UK: Univ. Pr. (1992) 418 p
  
\bibitem{nustar}
 T.~E.~Jeltema and S.~Profumo,
  arXiv:1108.1407 [astro-ph.HE].
  
\bibitem{colacoma}
S.~Colafrancesco, S.~Profumo and P.~Ullio,
  Astron.\ Astrophys.\  {\bf 455}, 21 (2006)
  [astro-ph/0507575].

\bibitem{dsphlim}
 M.~Ackermann {\it et al.}  [Fermi-LAT Collaboration],
  Phys.\ Rev.\ Lett.\  {\bf 107}, 241302 (2011)
  [arXiv:1108.3546 [astro-ph.HE]].


\bibitem{ullio}
L.~Bergstrom and P.~Ullio,
  Nucl.\ Phys.\ B {\bf 504}, 27 (1997)
  [hep-ph/9706232].


\bibitem{weniger}
 C.~Weniger,
  JCAP {\bf 1208}, 007 (2012)
  [arXiv:1204.2797 [hep-ph]];
 T.~Bringmann, X.~Huang, A.~Ibarra, S.~Vogl and C.~Weniger,
  JCAP {\bf 1207}, 054 (2012)
  [arXiv:1203.1312 [hep-ph]].

\bibitem{sufink}
  M.~Su and D.~P.~Finkbeiner,
  arXiv:1206.1616 [astro-ph.HE].
  


\end{thebibliography}

\section*{}
\clearpage
\addcontentsline{toc}{section}{Index}
\printindex

\end{document}